\documentclass[a4paper,twocolumn,showpacs
]{revtex4}

\usepackage{graphicx}
\usepackage{amssymb}
\usepackage{amsmath,amsfonts,latexsym}
\usepackage{dcolumn}               

\DeclareMathOperator{\tr}{tr}

\DeclareMathOperator{\Ry}{\mathcal{R}y}
\DeclareMathOperator*{\Simiq}{\simeq}

\DeclareMathOperator*{\To}{\to}

\newcommand{\vect}[1]{{\mathbf #1}}
\newcommand{\vectgr}[1]{{\boldsymbol#1}}    

\newcommand{\Frac}[2]{\displaystyle\frac{#1}{#2}}
\newcommand{\smc}[1]{\text{\sc{#1}}}

\begin{document}

\title{Condensation of Cavity Polaritons in a Disordered Environment}

\author{F. M. Marchetti${}^{1}$}

\author{B. D. Simons${}^{1}$}

\author{P. B. Littlewood${}^{1,2}$}

\affiliation{${}^{1}$Cavendish Laboratory, University of Cambridge,
             Madingley Road, Cambridge CB3 0HE, UK\\
             ${}^{2}$National High Magnetic Field
             Laboratory, Pulsed Field Facility, LANL, Los Alamos,
             NM87545} 
\date{Oct 27, 2004}       

\begin{abstract}
  A model for direct two band excitons in a disordered quantum well
  coupled to light in a cavity is investigated. In the limit in which
  the exciton density is high, we assess the impact of weak
  `pair-breaking' disorder on the feasibility of condensation of
  cavity polaritons. The mean-field phase diagram shows a `lower
  density' region, where the condensate is dominated by electronic
  excitations and where disorder tends to close the condensate and
  quench coherence. Increasing the density of excitations in the
  system, partially due to the screening of Coulomb interaction, the
  excitations contributing to the condensate become mainly photon-like
  and coherence is reestablished for any value of disorder. In
  contrast, in the photon dominated region of the phase diagram, the
  energy gap of the quasi-particle spectrum still closes when the
  disorder strength is increased. Above mean-field, thermal, quantum
  and fluctuations induced by disorder are considered and the spectrum
  of the collective excitations is evaluated. In particular, it is
  shown that the angle resolved photon intensity exhibits an abrupt
  change in its behaviour, going from the condensed to the
  non-condensed region.
\end{abstract}


\maketitle

\section{Introduction}
\label{sec:intro}
When the interaction between light and matter is strong, photons
propagating in a dielectric can couple to excitons and form composite
Bose particles known as polaritons~\cite{hopfield}. Since these
composite bosons have a light mass, there has long been interest in
the possibility that they might undergo Bose condensation at
relatively high transition temperatures. While the concept of a Bose
condensate of bulk polaritons has been discussed
extensively~\cite{keldysh,snoke_book}, in the open system, low-energy
polaritons are merely long wavelength photons, which are not
conserved~\cite{haug}. As a result, polaritons are unable to condense
into the ground state making the bulk polariton condensate an
intrinsically non-equilibrium phenomenon. However, providing the
lifetime of the polaritons is long compared to the thermalization
time, the confinement of the photons in a microcavity may allow for
the development of a quasi-equilibrium condensate.

In recent years, improvements in the technology of semiconductor
quantum wells has made the study of high-Q strongly-coupled planar
microcavities almost routine for III-V, II-VI and some organic
semiconductors~\cite{weisbuch,lesidang,lidzey}. Pumped both near
resonance~\cite{savvidis_baumberg_saba} (i.e. the cavity is excited at
a `critical' angle) or non-resonantly~\cite{lesidang,huang,deng}
(i.e. the system is excited at large angles), sharp superlinear
increase of the low energy polariton photoluminescence has been
observed.

Polaritons are characterised by a very short radiative lifetime (of
the order of $\tau_{\text{leak}} \sim 3$ ps) due to leakage from the
cavity. This has to be compared with the time scale to establish
equilibrium with the lattice, $\tau_{\text{lattice}}$, and the
equilibration time of the polaritons themselves,
$\tau_{\text{polariton}}$; the former controlled largely by
exciton-phonon scattering and the latter by exciton-exciton scattering
processes. Dynamical (or quasi-equilibrium) condensation of polaritons
can be reached when $\tau_{\text{polariton}} < \tau_{\text{leak}} <
\tau_{\text{lattice}}$. Due to the `bottleneck effect'~\cite{tassone},
thermalization of small momenta polaritons by acoustic phonons is
suppressed, causing a slow relaxation of polaritons in the low energy
state. This process is found to be strongly suppressed at higher
values of the pump power where superlinear emission
occurs~\cite{tartakovskii}. Recently, evidence of quasi-equilibrium
condensation, where the polariton life time reaches the relaxation
time, including second order coherence in the optical field, has been
observed~\cite{deng}.

Within a bosonic picture involving tightly bound excitons, much work
has been published on polariton dynamics under non-resonant
excitation, e.g. by making use of a rate equation
approach~\cite{pol_dynamics}. In this paper, we will focus on the
signatures of the fermionic nature of polaritons when the excitation
density increases. Here, the theoretical framework is not fully
established even in the case of thermal equilibrium. The relative role
of Coulomb forces (i.e. direct interaction between electron-hole
pairs) and photon mediated interactions has not been studied. There is
also an incomplete understanding of the effects of disorder, inelastic
scattering and `decoherence' phenomena, this being very important in
drawing a distinction between polariton condensation and conventional
lasing. Furthermore, previous theories, including those that
incorporate effects due to disorder and decoherence, have been at the
mean field level. Our aim here is to address all three of these
issues.

\subsection{Background}
\label{sec:backg}
Beginning with the pioneering work of Keldysh and
coworkers~\cite{keldysh_kopaev_kozlov}, the continuous
transition between the Bose condensed and BCS-like phase of the
equilibrium electron-hole system has been discussed within the
framework of a mean-field theory characterised by an order parameter
involving the coherent polarisation. Refinements of the mean-field
theory by Comte and Nozi\`eres~\cite{comte_nozieres} to include the
effects of screening provided a consistent theory of the electron-hole
liquid and condensate phases. In recent years, considerable efforts
have been made to explore the influence of light-matter interaction in
the non-equilibrium electron system driven by an external phase
coherent laser source --- the optical Stark
effect~\cite{stark}. However, only recently has the effect of quantum
phase coherent coupling of photons and excitons in the closed system
been addressed.

Cavity polaritons interact both directly via the Coulomb coupling of
their excitonic part, and indirectly via occupancy constraints on the
excitonic component, sometimes referred to as phase-space
filling. Since polariton masses are typically very small, the
coherence temperature increases rapidly with number density. Once the
characteristic interaction energy, either the dipole coupling energy
or the Coulomb interaction, is larger than the coherence temperature
of free polaritons, one must use an interacting theory. If the density
is still low enough that excitons are not strongly overlapping, an
appropriate approximation is to treat the excitons as bosons with a
short range repulsion, or equivalently as two level systems. However,
if the exciton density is high enough that the excitons become unbound
then the separate electron and hole (fermionic) degrees of freedom are
recovered.

Note then that there are at least two and generally three crossovers
as a function of increasing excitation: from a dilute gas of weakly
interacting polaritons to a strong coupling regime, where nevertheless
excitons are bound; then to a regime where the density of excitons is
high enough that they unbind (and which may be weakly coupled to
light, in the sense that the Rabi frequency is small in comparison to
the kinetic energy). If, as is usual, the latter regime is one of weak
coupling, nevertheless at high enough excitation one may reach a
strong coupling regime where the optical field provides the largest
scale in the problem.  It is in the intermediate, fermionic, regime of
weak coupling, that conventional semiconductor lasers
operate~\cite{laser_book}.

Previous theoretical work has addressed the question of the existence
of coherent, condensed polariton states in a simplified model. Taking
just a single cavity photon mode, polariton condensation has been
explored within the framework of an effective `zero-dimensional' model
in which excitons are trapped in isolated quantum
dots~\cite{paul}. Treated as two-level systems, the Hamiltonian for
the coupled exciton-photon system can be presented as a Dicke
model. The model allows for an inhomogeneous broadening of the levels,
while a quasi-equilibrium condition was imposed by fixing the total
number of excitations. At the level of mean-field (which becomes exact
when the number of exciton states coupled by the same microcavity
field is large), the ground state of the polariton system can be
written as a variational wave function. This wave function is a
superposition of coherent states of the exciton and photon in which
both the amplitude of the cavity field and the electron polarisation
acquire a non-zero expectation value.

Within the same two-level system model, the effect of decoherence
processes have recently been studied in~\cite{marzena}. Treated again
at the mean-field level, decoherence gradually suppresses the
excitonic component of the order parameter, driving the system toward
a gapless, weak-coupling regime similar to that of a conventional
laser.

Although the zero-dimensional system does provide intuition about the
collective properties of the polariton condensate, by its
construction, it is not able to discriminate between the BEC and
BCS-like character of the condensate phase in the low and high density
limits. Furthermore, the restriction of the model to a single cavity
photon mode does not allow for the consideration of fluctuations and
dispersion of the excitations in the polariton condensate
phase. Finally, in the two-level formulation, Coulomb interaction is
not considered self-consistently, and both at mean-field level and
beyond mean-field, the coherent polarisation and the coherent photon
field are locked together. Among other consequences, this means that
in this model the screening of Coulomb interaction by an increase of
the density of excitations cannot be taken into account. On the other
hand, such an effect is mimicked by the fermionic space filling
effect, which also has the result of increasing the photon component
at high excitation levels.

In the following, we will tune the strength of the light-matter
interaction so as to explore the interplay of exciton and polariton
condensation in a quasi two-dimensional microcavity system. Here, in
contrast to the localised exciton theory, one can expect weak
potential disorder to impact strongly on the integrity of the
condensate phase~\cite{zittartz}. Therefore, to keep our discussion
general, we will include the symmetry-breaking effect of a weak
potential disorder on the condensate. Although the mean-field theory
has the capacity to describe the entire continuous interpolation
between the low density BEC phase and the high density BCS-like phase,
for simplicity, we will limit our analysis of the mean-field theory
and fluctuations to the high density phase.

\subsection{Model}
\label{sec:model}
The many-body Hamiltonian for the coupled electron-hole/photon system
can be separated into constituent components according to
\begin{equation}
  \hat{\mathcal{H}} -\mu \hat{N}_{\text{ex}} =
  \hat{\mathcal{H}}_{\text{ei}} + \hat{\mathcal{H}}_{\text{dis}} + 
  \hat{\mathcal{H}}_{\text{ph}} + \hat{\mathcal{H}}_{\text{int}} \; .
\label{eq:model}
\end{equation}
Here, $\hat{\mathcal{H}}_{\text{ei}}$ represents the interacting
Hamiltonian of the electron and hole degrees of freedom in a
semi-conducting quantum well, while $\hat{\mathcal{H}}_{\text{dis}}$
describes the random potential generated by quenched impurities and
lattice defects.  $\hat{\mathcal{H}}_{\text{ph}}$ denotes the
Hamiltonian associated with the cavity photons and finally
$\hat{\mathcal{H}}_{\text{int}}$ represents the exciton-photon dipole
coupling interaction.

\begin{figure}
\begin{center}
\includegraphics[width=0.9\linewidth,angle=0]{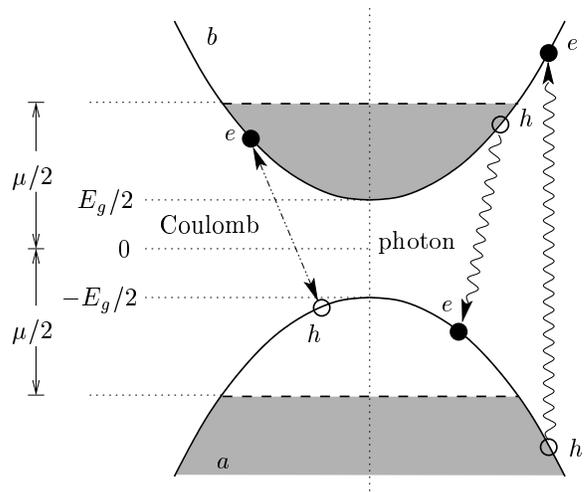}
\end{center}
\caption{\small Schematic picture of the valence ($a$) and conductance
	($b$) band, and the interactions included in the
	model~\eqref{eq:model}.} 
\label{fig:eband}
\end{figure}

Following the notation introduced in Ref.~\cite{zittartz} and adopting
an `electron picture' in which $b_{\vect{p}}$ and $b_{\vect{p}}^\dag$
($a_{\vect{p}}$ and $a_{\vect{p}}^\dag$) represent the annihilation
and creation operators of an electron with momentum $\vect{p}$ in the
conductance (valence) band (see Fig.~\ref{fig:eband}), the Hamiltonian
for a direct band-gap semiconductor is given by
\begin{multline*}
  \hat{\mathcal{H}}_{\text{ei}} = \sum_{\vect{p}} \xi_{\vect{p}}
  \left(b_{\vect{p}}^\dag b_{\vect{p}} + a_{\vect{p}}
  a_{\vect{p}}^\dag\right) \\
  + \Frac{1}{2} \sum_{\vect{q} \ne 0} v(\vect{q})
  \left(\rho_{\vect{q}} \rho_{-\vect{q}} - \sum_{\vect{p}}
  b_{\vect{p}}^\dag b_{\vect{p}} - \sum_{\vect{p}} a_{\vect{p}}
  a_{\vect{p}}^\dag\right) \; ,
\end{multline*}
where $\rho_{\vect{q}} = \sum_{\vect{p}} (b_{\vect{p} + \vect{q}}^\dag
b_{\vect{p}} - a_{\vect{p}} a_{\vect{p} + \vect{q}}^\dag)$ is the
total electron density operator. Here, for simplicity, we suppose that
the electrons are spinless. Taking the middle of the gap as the energy
reference, we assume a parabolic conduction and valence band of
particles
\begin{equation}
  \xi_{\vect{p}} = \Frac{\vect{p}^2}{2m} - \Frac{\mu - E_g}{2} 
  \equiv\Frac{\vect{p}^2}{2m} - \varepsilon_F
  \; ,
\label{eq:parab}
\end{equation}
where, for simplicity, we have assumed the electrons and holes exhibit
the same effective mass, $m$. At the level of mean-field, one can
confirm that this assumption is innocuous and corresponds to replacing
the effective mass with the double of the reduced mass. In contrast,
deviation from parabolicity in the physical system impairs the
capacity for perfect nesting of the Fermi surface and reduces the
potential for pair correlation. Finally, for convenience, we have
introduced in Eq.~\eqref{eq:parab} the concept of an effective Fermi
energy $\varepsilon_F\equiv p_F^2/2m=(\mu- E_g)/2$, which, along with
the chemical potential $\mu$, is determined self-consistently as a
non-trivial function of the total number of excitations in the
system~\eqref{eq:nexci} (see later in section~\ref{sec:chemi}).

In addition to the lattice potential, the electrons and holes
experience a Coulomb interaction. In the \emph{high density regime}
$\rho_{el} a_0^2 \gg 1$, where $\rho_{el} = p_F^2/4\pi$ denotes the
areal density of the particles, (i.e. when the the excitonic Bohr
radius $a_0 = 2 \epsilon_0 /e^2 m$ is much larger than the average
distance between electron and holes $1/\sqrt{\rho_{el}}$), the Coulomb
interaction is screened due to both electrons and
holes~\cite{zittartz,comte_nozieres}:
\begin{equation*}
  v(\vect{q}) = \Frac{4 \pi e^2}{\epsilon_0(\vect{q}^2 + \kappa^2)} \;
.
\end{equation*}
In the two-dimensional system, the screening length is approximately
set by the Bohr radius $1/\kappa=a_0$. Hence, in this limit, the
Coulomb interaction can be replaced by a short range contact
interaction~\cite{zittartz},
\begin{equation*}
  \sum_{\vect{p}'} v (\vect{p} - \vect{p}') \Big|_{|\vect{p}'|=p_F}
  \simeq
  \begin{cases}
    \bar{V} & |p_F-p| < \kappa\\
    0       & |p_F-p| > \kappa\; ,
  \end{cases}
\end{equation*}
with a strength given by the angular average over the Fermi surface
\begin{equation}
  \bar{V} = \Ry a_0^3 \Frac{8 \pi}{\sqrt{1 + (2p_F a_0)^2}}\; .
\label{eq:ccons}
\end{equation}
Here $\Ry =e^2/2\epsilon_0 a_0$ denotes the Rydberg energy associated
with the exciton and sets the characteristic energy scale of the
interaction. Therefore, as a consequence of the screening, an increase
in the total number of electronic excitations leads to a reduction in
the effective strength of the Coulomb interaction. Notice that,
although the condensed state may have a gap, we shall be discussing
the dense limit, where the coherence length is much longer than
$1/\kappa$, and so this will not influence the important short range
part of the potential.

To incorporate the effect of disorder, we suppose that the
electron-hole system is subject to a generic weak impurity potential,
\begin{equation*}
  \hat{\mathcal{H}}_{\text{dis}} = \int d\vect{r}\; \left[U_{b}
  (\vect{r}) b^\dag (\vect{r}) b (\vect{r}) + U_{a} (\vect{r}) a^\dag
  (\vect{r}) a (\vect{r})\right] \; .
\end{equation*}
Naively, since the electrons and holes carry opposite charge, their
response to potential impurities should be equal in magnitude and
opposite in sign. However, differences in the dielectric properties of
the conduction and valence band electrons, combined with the potential
asymmetry imposed by the quasi two-dimensional geometry, implies that
the potentials $U_{a}$ and $U_{b}$ are not perfectly
anticorrelated. Furthermore, there are mechanisms of disorder such as
variations in the width of the quantum well through atomic terracing,
strain fields, etc., that act on electrons and holes with the same
sign. Therefore, in the following, we will suppose that a generic
disorder potential is comprised of two channels: A potential
$U_n(\vect{r}) \equiv [U_b(\vect{r}) - U_a(\vect{r})]/2$ which acts
symmetrically as a `charge-neutral' potential and, as we will see,
does not effect the integrity of the condensate phase. A second,
statistically independent, potential $U_c(\vect{r}) \equiv
[U_b(\vect{r}) + U_a(\vect{r})]/2$ acts as an asymmetric `charged'
component, and presents a pair-breaking perturbation which gradually
destroys the excitonic condensate. As mentioned above, its effect in
the dense exciton system on the integrity of the exciton insulator
phase has been explored in an early work by
Zittartz~\cite{zittartz}. In both cases, we will suppose that the
potentials $U_{n,c}$ are drawn at random from a Gaussian white-noise
distribution, with zero mean, and variance given by
\begin{equation}
  \langle U_{n,c} (\vect{r}_1) U_{n,c} (\vect{r}_2)\rangle =
  \Frac{1}{2\pi \nu\tau_{n,c}} \delta(\vect{r}_1 - \vect{r}_2) \; ,
\label{eq:disor}
\end{equation}
where $\tau_n$ and $\tau_c$ are the associated scattering times, and
$\nu = m/2\pi$ is the two-dimensional density of states of the
spinless electron system. Restricting attention to the high density
phase, setting $\varepsilon_{F} \tau_{n,c} \gg 1$, we will further
suppose that the impurity potential impose only a weak perturbation on
the electron and hole degrees of freedom, allowing therefore the
dynamics to be treated quasi-classically. This completes the
description of the electron-hole system.

Free photons in the cavity are described by the microscopic quasi
two-dimensional Hamiltonian
\begin{equation*}
  \hat{\mathcal{H}}_{\text{ph}} = \sum_{\vect{p}}
  \psi_{\vect{p}}^\dag\left[\omega (\vect{p}) - \mu\right]
  \psi_{\vect{p}} \; ,
\end{equation*}
where their dispersion, $\omega (\vect{p}) = \sqrt{\omega_c^2 +
(c\vect{p})^2}$, is quantised in the direction perpendicular to the
plane of the cavity mirrors, $\omega_c = c\pi j/L_{\text{m}}$, where
$L_{\text{m}}$ denotes the distance between the mirrors, and $j\in
\mathbb{Z}$.  In the following, we will suppose that the electron-hole
system engages just a single sub-band for which $\omega_c\ne 0$. In
practice, providing the sub-band separation is substantially larger
than the quasi-particle energy gap that develops in the electron-hole
system, neighboring sub-bands can be safely neglected.

Finally, in the dipole or `rotating-wave' approximation, the photons
are assumed to be coupled to the electron-hole system through a local
interaction,
\begin{equation*}
  \hat{\mathcal{H}}_{\text{int}} = g \int d\vect{r}\left[\psi
  (\vect{r}) b^\dag (\vect{r}) a (\vect{r}) + \text{h.c.}\right] \; .
\end{equation*}
Here, terms which do not conserve the number of excitations (i.e. the
ones describing spontaneous creation or annihilation of a photon and
an exciton) have been neglected. On one hand, we expect this
approximation to be valid in the weak coupling limit between photon
and matter, where these additional terms can be `integrated out',
giving a small `renormalisation' of the single particle electronic and
photonic dispersion laws. At the same time, since we are dealing with
a system where the temperature is much smaller than the photon
frequency, we can neglect the tiny spontaneous population that would
be generated by these non-resonant terms.

To mimic the effect of the external photon source, we suppose that the
electron-hole/photon system is held in quasi-equilibrium by tuning the
chemical potential $\mu$ in~\eqref{eq:model} to fix the total number
of excitations
\begin{equation}
  \hat{N}_{\text{ex}} = \sum_{\vect{p}} \psi^\dag_{\vect{p}}
  \psi_{\vect{p}} + \Frac{1}{2} \sum_{\vect{p}}
  \left(b^\dag_{\vect{p}} b_{\vect{p}} + a_{\vect{p}}
  a^\dag_{\vect{p}}\right)\; .
\label{eq:nexci}
\end{equation}
However, how the system chooses to portion the excitations between the
electron-hole and photon degrees of freedom depends sensitively on the
properties of the condensate.

This completes the formal construction of the microscopic many-body
Hamiltonian. In order to explore the mean-field content and the
collective excitations of the light-matter system, we will draw on the
theory of the weakly disordered superconductor. Using conventional
field theoretic methods, we will show that the low-energy properties
of the system can be cast in the framework of an effective field
theory with an action of non-linear sigma-model
type~\cite{efetov}. Partitioning our discussion into properties of the
mean-field content of the theory and the role of fluctuations, we will
focus on the phase diagram of the system as a function of density and
disorder on one side, and the nature of the collective excitations on
the other. Since the field theory technology disguises much of the
physical content, we will close the introduction with an overview of
the findings of the mean-field theory. As well as summarising the main
results, it will provide a useful prospective for the formal analysis.

\begin{figure}
\begin{center}
\includegraphics[width=1\linewidth,angle=0]{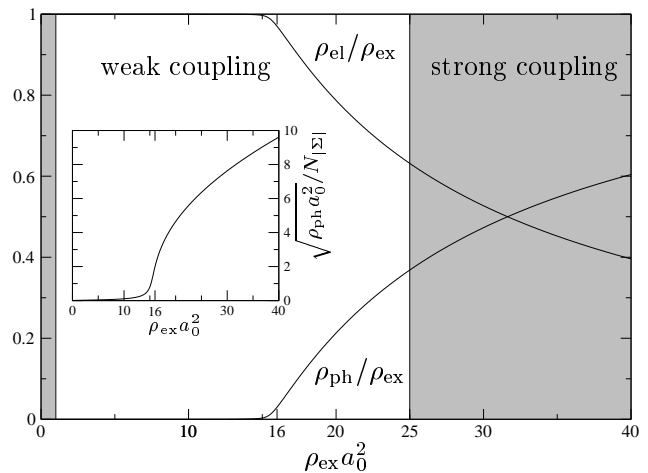}
\end{center}
\caption{\small Relative number of (total) electrons $n_{\text{el}} =
        \rho_{\text{el}}/ \rho_{\text{ex}}$ and (condensed) photons
        $n_{\text{ph}} = \rho_{\text{ph}}/ \rho_{\text{ex}}$ are
        plotted against $\rho_{\text{ex}} a_0^2$ for fixed values of
        the coupling constants $\tilde{g} = 1$ and $\lambda_c = 10$
        and for $d = 200$. Note that, for $\rho_{\text{ex}} a_0^2 >
        25$, the system enters a strong coupling regime, where our
        model cannot be applied. In the inset, the number of photons
        $\rho_{\text{ph}}a_0^2$ is compared with the number of
        electrons in the condensate, $N_{|\Sigma|} = |\Sigma| \nu
        a_0^2$ (see Eq.~\eqref{eq:intoc}). For $\rho_{\text{ex}} a_0^2
        > d/4\pi \simeq 16$, the excitations in the condensate are
        dominated by photons.}
\label{fig:nehph}
\end{figure}

\begin{figure}
\begin{center}
\includegraphics[width=1\linewidth,angle=0]{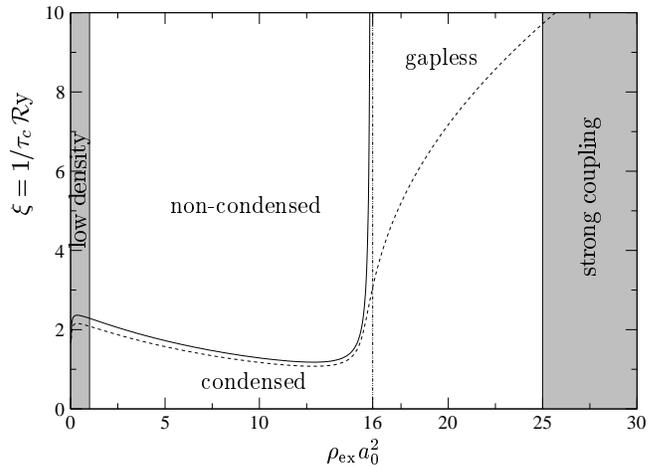}
\end{center}
\caption{\small Mean-field phase diagram for $\tilde{g} = 1$,
        $\lambda_c = 10$ and $d = 200$; $\rho_{\text{ex}} a_0^2$
        represents the excitation density, while $\xi = 1/\tau_c \Ry$
        the disorder strength.}
\label{fig:phase}
\end{figure}

In the high density regime, the mean-field content of the action will
be shown in section~\ref{sec:meanf} to be insensitive to the `neutral'
component of the disorder potential, $U_n (\vect{r})$, a manifestation
of the Anderson theorem in the exciton context. This reproduces the
mean-field thermodynamic properties of a system subject only to a
`charged' weakly disorder potential, $U_c (\vect{r})$. By analysing
these mean-field equations, the generalisation of those derived by
Zittartz~\cite{zittartz} for the exciton insulator system, we will
describe the rearrangement of the ground state due to condensate
formation. In particular, we will show that the condensate of the
polariton system is characterised by an order parameter which engages
the combination of the phase locked polarisation and photon
amplitudes. The mean-field phase diagram (see Fig.~\ref{fig:phase})
can be characterised in terms of the total density of excitations,
$\rho_{\text{ex}} a_0^2$, the disorder strength $\xi = 1/\tau_c \Ry$
and the following dimensionless material parameters, respectively the
coupling to the photon field and the Coulomb coupling strength (see
expression~\eqref{eq:ccons}):
\begin{equation}
\begin{split}
  \tilde{g} &= g\sqrt{\Frac{\nu L^2}{\Ry}} \\ 
  \tilde{g}_c(x) &= g_c \nu L^2 = \Frac{\lambda_c}{\sqrt{1 +
  4(d-x)}} \; .
\label{eq:resca}
\end{split}
\end{equation}
Here, the constant of proportionality $\lambda_c = 4a_0/L_\perp$
depends on the thickness $L_\perp$ of the quantum well, $x = (\omega_c
- \mu)/\Ry$ denotes the characteristic energy separation between the
photon band edge and the chemical potential (measured in units of the
Rydberg), and $d = (\omega_c - E_g)/\Ry > 0$. Note that, as in BCS
theory, $g_c$ scales with the volume, i.e. in the thermodynamic limit
$g_c L^2$ is finite while, by contrast, $g$ scales with the square
root of the volume, i.e. $g^2 L^2$ is finite.

When the density $\rho_{\text{ex}} a_0^2$ is low, the majority of
excitations are invested in electrons and holes (see
figure~\ref{fig:nehph}) and therefore, as the density increases, the
chemical potential rises linearly. Here, providing the density is
large enough to keep the particles unbound, the condensed phase is
reminiscent of the exciton insulator one, even if a small fraction of
photons do contribute to the condensate (see the inset of
figure~\ref{fig:nehph}). Eventually, as the chemical potential
approaches the band edge, $\omega_c$, the photons are brought into
resonance and the character of the condensate changes abruptly. Here,
the excitations become increasingly photon-like, with
$\rho_{\text{ex}} a_0^2$ diverging exponentially when the chemical
potential converges on $\omega_c$. At the same time, screening
suppresses the excitonic coupling constant, determining the majority
of the excitation in the condensate to acquire a photon-like
character. This is clearly shown in figure~\ref{fig:nehph}, where the
normalised number of \emph{total} electrons $\rho_{\text{el}}/
\rho_{\text{ex}}$ and photons $\rho_{\text{ph}}/ \rho_{\text{ex}}$
(see Eq.~\eqref{eq:rscle}) are plotted as a function of the density of
excitations. In the inset of the picture, instead, the fraction of
photons to excitons \emph{in the condensate} (for the exact
definition, see later on, Eq.~\eqref{eq:intoc}) clearly shows an
abrupt growth as the density of excitations in increased above the
value $d/4\pi$, which we will see to coincide with the maximum density
which can be reached at $\mu = \omega_c$ in absence of photons.

Staying at the level of mean-field, the `charged' component of the
disorder potential presents a symmetry breaking perturbation which
depletes the excitonic component of the condensate. As such, its
effect on the condensate depends sensitively on the density of
excitations. When $1<\rho_{\text{ex}} a_0^2 < d/4\pi$, the condensate
has a particle-hole character and the phase diagram
(Fig.~\ref{fig:phase}) mimics the behaviour of the symmetry broken
exciton insulator~\cite{zittartz} -- and therefore of a
superconducting system in presence of magnetic impurities. When the
scattering rate $\tau_c$ is comparable to the value of the unperturbed
order parameter, the system enters a gapless phase before the
condensate is extinguished altogether. At larger densities, the
condensate becomes photon-dominated and therefore robust against any
value of the disorder potential. Interestingly, however, the residual
effect of the disorder exposes a substantial region of the phase
diagram where the system is gapless. Here, the condensate acquires all
of the conventional characteristics of a semiconductor laser ---
i.e. a substantial coherent optical field, but a gapless spectrum of
electron-hole pairs with negligible electronic
polarisation~\cite{sclaser}. Once again, at sufficiently large
densities, the system enters a strong coupling regime, where the
validity of the model becomes doubtful.

This completes our summary of the mean-field content of the
theory. The remainder of the paper is organised as follows. In the
next section, we develop a field theory of the weakly disordered
polariton system. Focusing on the saddle-point approximation, in
section~\ref{sec:meanf} we elaborate and expand our discussion of the
mean-field content of the model. Using these results as a platform and
exposing analogies with the symmetry broken superconducting system, in
section~\ref{sec:excit} the attention is then turned to the impact of
thermal, quantum and disordered driven fluctuations. Here, we explore
the spectrum of the collective excitations and, in particular, we
discuss how the transition to the condensed phase provide a signature
in the photoluminescence. Finally, in section~\ref{sec:concl}, we
conclude.

\section{Path Integral Formulation}
\label{sec:cpath}
The development of a low-energy field theory of the weakly disordered
exciton-light system mirrors closely the quasi-classical theory of the
disordered superconductor. By presenting the quantum partition
function of the many-body system as a coherent state path integral,
the ensemble or impurity averaged free energy can be cast in terms of
a replicated theory. At the level of the mean-field theory, this
approach is equivalent to the self-consistent Hartree-Fock treatment
where the impact of disorder is treated in the self-consistent Born
approximation. Using the mean-field result as a platform, the field
theoretic formulation provides the means to develop a field theory of
the low-energy excitations of the polariton condensate.

The quantum partition function $\mathcal{Z} = \tr [e^{-\beta
(\hat{\mathcal{H}} -\mu \hat{N}_{\text{ex}})}]$, where $\beta =
1/k_{\smc{b}} T$ is the inverse temperature, can be expressed as a
coherent state path integral over fermionic and bosonic fields. To
facilitative impurity averaging of the \emph{free energy} over the
disorder potentials~\eqref{eq:disor}, it is convenient to engage the
`replica trick'~\cite{edwards_anderson},
\begin{equation*}
  F = - \Frac{1}{\beta} \langle \ln \mathcal{Z}\rangle = -
  \Frac{1}{\beta} \lim_{n\to 0} \Frac{\langle \mathcal{Z}^n \rangle
  -1}{n} \; .
\end{equation*}
Although the analytic continuation presents difficulties in treating
non-perturbative structures, the technique can be applied safely to
the mean-field and collective properties discussed in this work. Once
replicated, a Hubbard-Stratonovich decoupling of the Coulomb
interaction obtains the field integral for the quantum partition
function
\begin{multline}
  \mathcal{Z}^n = \int D(\psi^* , \psi) \int D(\Sigma^* , \Sigma)\;
  \int D(\phi^\dag , \phi)\, e^{-S_\text{ph} - S_{\Sigma}} \\
  \exp\left[-\int_0^\beta d\tau \int d\vect{r} \; \phi^\dag (\vect{r}
  , \tau) \left(\partial_\tau + \hat{H}\right)\phi (\vect{r} ,
  \tau)\right] \; ,
\label{eq:parti}
\end{multline}
where, in order to lighten the notation, we have omitted the replica
index carried by all of the fields. Here, $S_\text{ph}$ and
$S_{\Sigma}$ correspond respectively to the bare photonic and
excitonic components of the action,
\begin{align}
\label{eq:phact}
  S_\text{ph} &= \int_0^\beta d\tau \sum_{\vect{p}}
  \psi_{\vect{p}}^* (\tau) \left[\partial_\tau + \omega (\vect{p})-
  \mu \right]\psi_{\vect{p}} (\tau)\\
  S_{\Sigma} &= \Frac{1}{g_c L^2} \int_0^\beta d\tau \int d\vect{r} \;
  |\Sigma (\vect{r} , \tau)|^2\; ,
\label{eq:eiact}
\end{align}
while, arranging the fermionic fields into a Nambu-like spinor
$\phi^\dag = (b^\dag , a^\dag)$ and $\phi = (b,a)^{\mathsf{T}}$, the
(single-particle) action for the internal electronic degrees of
freedom (which, with an abuse of terminology, we will later on refer
to as `particle-hole' components) takes the form reminiscent of the
Gor'kov or Bogoliubov de-Gennes superconducting Hamiltonian,
\begin{equation}
  \hat{H} = \left[\hat{\xi}_{\hat{\vect{p}}} + U_n (\vect{r})\right]
  \sigma_3 + U_c(\vect{r}) + \widehat{\Delta}\; .
\label{eq:singl}
\end{equation}
Here, $\sigma_i$ represent Pauli matrices operating in the
particle-hole subspace, and $\Delta = |\Delta| e^{i\chi}$ represents
the complex `composite order parameter'
\begin{equation*}
  \widehat{\Delta} = \begin{pmatrix}
    0 & \Sigma + g\psi\\
    \Sigma^* + g\psi^* & 0
  \end{pmatrix} = |\Delta| \sigma_1 e^{- i \chi \sigma_3}\; .
\end{equation*}
involving both the polarisation $\Sigma$ and photon $\psi$ fields.
Taking into account the contact nature of the screened Coulomb
interaction, the coupling constant $g_c$ can be derived starting from
the expression~\eqref{eq:ccons} and is given in
Eq.~\eqref{eq:resca}. At the level of mean-field, the excitonic order
parameter signals the development of the anomalous pairing amplitude
\begin{equation*}
  \Sigma (\vect{r})= - g_c L^2 \langle \text{g.s.}| b^\dag
  (\vect{r}) a (\vect{r})|\text{g.s.} \rangle \; ,
\end{equation*}
while the ordinary Hartree-Fock pairings, $\langle \text{g.s.}| b^\dag
(\vect{r}) b (\vect{r})|\text{g.s.}  \rangle$ and $\langle
\text{g.s.}| a (\vect{r}) a^\dag (\vect{r})|\text{g.s.} \rangle$,
crucial in the low-density limit, effect only a small renormalisation
of the single-particle energy $\xi_{\vect{p}}$.  Restricting attention
to the high density regime, such renormalisations can be neglected (or
absorbed into a redefinition of the position of the band edge).

As noted above, the single-particle Hamiltonian for the electron and
hole mirrors the structure of the quasi-particle Hamiltonian for a
superconductor. In particular, in the absence of disorder, an electron
of energy $\xi_{\vect{p}}$ in the valance band can pair with a hole of
energy $\xi_{\vect{p}}$ in the conduction band. The development of a
homogeneous order parameter $\Delta$ induces a gap in the
quasi-particle density of states at the effective Fermi level
$\varepsilon_F$. Already at this level, the insensitivity of the
condensate to `neutral' disorder $U_n$ is apparent: Even in the
presence of the disorder potential, the electron with a
single-particle energy can pair with a partner hole with the same
energy and form a homogeneous condensate, a reflection of the Anderson
theorem of superconductivity in the electron-hole system.  By
contrast, the symmetry breaking potential $U_c$ lifts the degeneracy
of the electron and hole degrees of freedom and imposes a
pair-breaking perturbation on the condensate. As a result, one expects
a gradual depletion of the condensate. In fact, we will see later on
in section~\ref{sec:phase}, when establishing the phase diagram, that,
since photons are almost insensitive to the effect of disorder, only
the `excitonic contribution' to the condensate will be affected, while
at very high densities, the photonic component will be able to restore
coherence.

In principle, one could explore the mean-field content of the coupled
system by developing a diagrammatic self-consistent Hartree-Fock
approximation. However, in the following, we will be concerned with
the impact of fluctuations on the integrity of the condensate
phase. As well as thermal and quantum fluctuations, mechanisms of
quantum interference affect the spectrum of quasi-particle excitations
above the condensate. To properly accommodate these effects, it is
helpful to construct an effective low-energy theory which encodes the
diffusive character of the dynamics.  To classify the different
channels of interference, it is helpful to identify the symmetry
properties of the coupled system.

As well as its invariance under the global gauge transformation
\begin{align}
  \nonumber \phi &\mapsto e^{i\chi' \sigma_3/2} \phi & \phi^\dag
  &\mapsto \phi^\dag e^{-i \chi' \sigma_3/2} \\ 
  \Sigma &\mapsto \Sigma\, e^{i\chi'} & \psi &\mapsto \psi
  e^{i\chi'}\; ,
\label{eq:gauge}
\end{align}
in the absence of the symmetry breaking potential $U_c$, the matrix
Hamiltonian~\eqref{eq:singl} exhibits the discrete particle-hole
symmetry $\hat{H} = - \sigma_2 \hat{H}^\mathsf{T} \sigma_2$. Its
presence facilitates mechanisms of quantum interference which effect
the low-energy properties of the quasi-particles
excitations. Following a standard procedure (see, e.g.,
Ref.~\cite{damian}), in order to properly take into account for the
`soft diffusion modes' associated with this symmetry, it is useful to
double the field space setting
\begin{align*}
  \Phi &= \Frac{1}{\sqrt{2}}
  \begin{pmatrix} \phi \\
  \sigma_2 {\phi^\dag}^{\mathsf{T}}
  \end{pmatrix}_{\smc{cc}} &
  \bar{\Phi} &= \frac{1}{\sqrt{2}} 
  \begin{pmatrix} \phi^\dag & -\phi^{\mathsf{T}} \sigma_2  
  \end{pmatrix}_{\smc{cc}} \; .
\end{align*}
As a result, the fields $\Phi$ engage a total of $4\times n$
components (i.e.  $n$ replica, particle-hole and `charge conjugation'
\smc{cc}), while the redundancy implied by the field doubling imposes
the symmetry relations $\Phi = \sigma_2 \gamma
\bar{\Phi}^{\mathsf{T}}$ and $\bar{\Phi} = -\Phi^{\mathsf{T}} \sigma_2
\gamma^{\mathsf{T}}$, with $\gamma = - i \sigma_2^{\smc{cc}}$. It is
important to note that the impact of the full charge-conjugation
structure of the fields is manifest only in the mesoscopic fluctuation
phenomena which effect only the low-energy quasi-particle
excitations. In particular, such mechanisms of quantum interference do
not effect the properties of the system at the level of the
saddle-point or mean-field.

Cast in the form~\eqref{eq:parti}, the action mirrors closely that
studied in the context of the disordered superconductor and, to
identify the effective low-energy theory, one may draw on the existing
literature. In particular, in the quasi-classical and dirty limits,
\begin{equation}
  \varepsilon_F \gg \Frac{1}{\tau_n} \gg \Frac{1}{\tau_c} , |\Delta|
  \; ,
\label{eq:scale}
\end{equation}
it has been shown~\cite{damian,lamacraft} that the long-range
properties of the system can be expressed in terms of an action of
nonlinear $\sigma$-model type,
\begin{equation*}
  \langle\mathcal{Z}^n\rangle =\int D(\psi^* , \psi) \int D(\Sigma^* ,
  \Sigma) e^{-S_\text{ph} - S_{\Sigma}} \int DQ\; e^{-S[Q]} \; ,
\end{equation*}
where
\begin{multline}
  S[Q] = \Frac{\pi \nu}{8} \int d\vect{r} \tr \left[D (\nabla Q)^2
  \right.\\ 
  \left.  - 4\left(\hat{\epsilon}\sigma_3 \sigma_3^\smc{cc} -
  i\widehat{\Delta}\sigma_3\right)Q - \Frac{1}{\tau_c} \left(\sigma_3
  \sigma_3^{\smc{cc}} Q\right)^2\right]\; .
\label{eq:actio}
\end{multline}
Here, $D = v_F^2 \tau_n/2$ represents the diffusion constant
associated with the `neutral' disorder, while the matrix field
$Q(\vect{r})= T(\vect{r}) \Lambda \sigma_3 \sigma_3^{\smc{cc}}
T^{-1}(\vect{r})$ carries replica, time and internal (particle-hole
and \smc{cc}) indices, and the generators $T$ are compatible with the
`charge-conjugation 'symmetry properties which $Q$ inherits from the
dyadic product $\sigma_3 \Phi \otimes \bar{\Phi}$. Qualitatively, the
fluctuations of the matrix field Q encode the effect of long-range
mechanisms of quantum interference.

Combined with the bare actions for the auxiliary
fields~\eqref{eq:phact} and~\eqref{eq:eiact}, the non-linear
$\sigma$-model action fully encodes the effect of weak disorder in the
cavity polariton system. In the absence of the order parameter, the
action describes the diffusive density relaxation of, separately, the
electron and hole degrees of freedom. The development of the order
parameter $|\Delta|$ describes the rearrangement of the ground state
due to the formation of the polariton condensate. Finally, scattering
in the channel of the `charged' disorder potential projects out the
diffusion modes which act coherently on the particle-hole degrees of
freedom. In order to explore the capacity of the system to undergo
condensation, it is necessary to investigate how the order parameter,
and symmetry breaking impurity potential revise the structure of the
saddle-point solution.

\section{Saddle-Point Analysis: Mean-Field Theory}
\label{sec:meanf}
The mean-field content of the theory is encoded in the saddle-point
structure of the effective action. Varying the total
action~\eqref{eq:actio} with respect to $Q$, subject to the nonlinear
constraint $Q^2 = \openone$, the saddle-point equation takes the form
\begin{multline}
  D \nabla (Q \nabla Q) + [Q , \hat{\epsilon}\sigma_3
  \sigma_3^\smc{cc} - i \widehat{\Delta}\sigma_3] \\ 
  + \Frac{1}{2 \tau_c} [Q, \sigma_3 \sigma_3^\smc{cc} Q \sigma_3
  \sigma_3^\smc{cc}] =0 \; .
\label{eq:motio}
\end{multline}
Such equations of motion are familiar from the quasi-classical theory
of the weakly disordered superconductor, where the field $Q$ is
identified as the average quasi-classical Green
function~\cite{usadel}. The second term describes the `rotation' of
the average quasi-classical Green function due to the formation of a
condensate $\Delta$, while the last term describes the pair-breaking
effect of the perturbation imposed by the symmetry breaking
disorder. 

Similarly, varying the action with respect to the excitonic order
parameter field $\Sigma$, and the photonic field $\psi$, one obtains
the coupled self-consistency equations:
\begin{multline}
  \Frac{1}{g_c} \Sigma_{\omega_h} (\vect{r}) = -\Frac{i \pi \nu
  L^2}{4\beta} \sum_{\epsilon_n} \tr \left[\sigma_- Q_{n+h,n}
  (\vect{r})\right]\\ 
  = \Frac{1}{g} \sum_{\vect{p}} e^{i \vect{p} \cdot \vect{r}}
  \left[i\omega_h + \omega (\vect{p}) - \mu\right]
  \psi_{\omega_h,\vect{p}}\; ,
\label{eq:selfo}
\end{multline}
where the matrix $\sigma_- = \sigma_1 - i\sigma_2$ projects onto the
off-diagonal particle-hole channel and where $\omega_h = 2l \pi/\beta$
are the bosonic Matsubara frequencies. Since the photonic and
excitonic interactions are decoupled in the same channel, it is not
surprising that, at the level of the saddle-point, the two order
parameters cannot be independent but instead are coupled together by
the constraint implied by~\eqref{eq:selfo}. This means that in the
condensed phase both order parameters develop in concert. In other
words, in the condensed phase of the present microscopic theory, both
electrons and photons participate in the formation of the condensate
--- it is only their relative proportion that depends (sensitively) on
the total number of excitations $N_{\text{ex}}$ present in the system.

\begin{figure}
\begin{center}
\includegraphics[width=1\linewidth,angle=0]{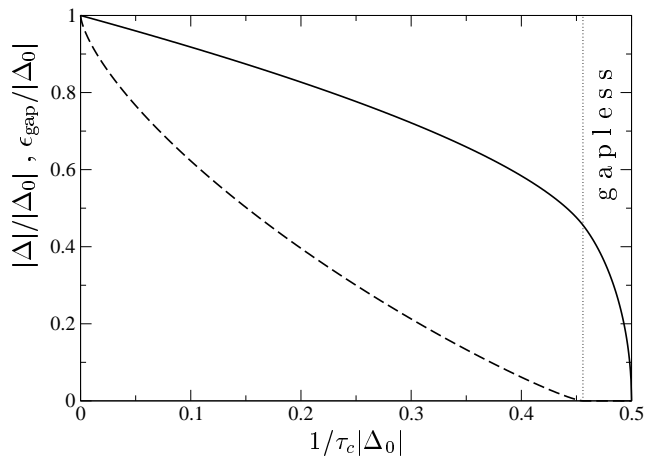}
\end{center}
\caption{\small The order parameter $|\Delta|/|\Delta_0|$ (solid line)
        and energy gap $\epsilon_{\text{gap}} / |\Delta_0|$ (dashed
        line) versus the strength of disorder $1/\tau_c
        |\Delta_0|$. Here, $|\Delta_0|$ represents the value of the
        order parameter in absence of disorder. The critical value at
        which the order parameter disappear is given by $1/\tau_c
        |\Delta_0| = 1/2$, while the gapless region starts at
        $1/\tau_c |\Delta_0| = e^{-\pi/4} \simeq 0.46$ (or
        equivalently at $\zeta = 1/\tau_c |\Delta| = 1$).}
\label{fig:engap}
\end{figure}

The mean-field equations~\eqref{eq:motio} and~\eqref{eq:selfo} can be
solved simultaneously with the mean-field Ansatz that the two order
parameters are space and time independent (i.e., $\Sigma
(\vect{r},\tau) = \Sigma$ and $\psi_{\omega_h,\vect{p}} = \psi
\delta_{\omega_h,0} \delta_{\vect{p},0}$), and the matrix field $Q
(\vect{r})$ is homogeneous in space and diagonal in the Matsubara
indices. Now, through the transformation $Q \mapsto e^{i\chi
\sigma_3/2} Q e^{-i\chi \sigma_3/2}$, the (homogeneous) global phase
$\chi$ of the composite order parameter $\Delta$ can be gauged out of
the action~\eqref{eq:actio} and the equation of
motion~\eqref{eq:motio} can be solved, giving
\begin{equation*}
  \bar{Q} = \Frac{u}{\sqrt{1 + u^2}}\; \sigma_3 \sigma_3^{\smc{cc}} -
  \Frac{1}{\sqrt{1 + u^2}}\; \sigma_2 \; ,
\end{equation*}
where the variable $u$ satisfies the equation~\cite{abrikosov_gorkov}:
\begin{equation}
  \Frac{\epsilon_n}{|\Delta|} = u - \Frac{1}{|\Delta|
  \tau_c}\Frac{u}{\sqrt{1 + u^2}} \; .
\label{eq:homog}
\end{equation}
As expected, in the mean-field approximation, the solution of the
quasi-classical equations of motion become insensitive to the symmetry
preserving disorder potential (i.e. the solution is independent of the
classical diffusion constant $D$). Such independence is a
manifestation of the Anderson theorem that protects the integrity of
the condensate to impurities which do not lift the particle-hole
symmetry of the system.

In the homogeneous case, the self-consistency equations for the order
parameters~\eqref{eq:selfo} require the phases of the excitonic order
parameter $\Sigma$ and of the photonic field $\psi$ to coincide, so
that $|\Delta| = |\Sigma| + g|\psi|$ and from which follows the
constraint:
\begin{equation}
  \Frac{1}{g} (\omega_c - \mu) |\psi| = \Frac{1}{g_c}|\Sigma|\; .
\label{eq:const}
\end{equation}
Indeed, the same restriction has been found in
Refs.~\cite{paul,marzena} in the context of the Dicke model for the
localised exciton-photon system.  While the chemical potential $\mu$
does not exceed the cavity mode edge $\omega_c$, an order parameter
$|\Delta| \ne 0$ is developed, to which both photons and the coherent
polarisation contribute. Once cast in terms of the effective order
parameter $\Delta$, the remaining saddle-point or mean-field equations
assume the canonical form first reported by Abrikosov and Gor'kov in
the context of the symmetry broken disordered
superconductor~\cite{abrikosov_gorkov}, and later, in the present
context, applied to the symmetry broken exciton
insulator~\cite{zittartz}. To make this correspondence explicit, one
may define the composite order parameter through, say, the photon
condensate fraction, and an effective coupling constant as
\begin{align*}
  |\Delta| &= \left[1 + \Frac{g_c}{g^2} (\omega_c - \mu)\right]
   g|\psi| \\
   g_{\text{eff}} &= g_c + \Frac{g^2}{\omega_c - \mu} \; ,
\end{align*}
after which the self-consistency equation reads
\begin{equation}
  \Frac{|\Delta|}{g_{\text{eff}} \nu L^2} = \Frac{\pi}{\beta}
  \sum_{\epsilon_n} \Frac{1}{\sqrt{1 + u^2}}\; .
\label{eq:self2}
\end{equation}
In the absence of symmetry-breaking disorder, $u=\epsilon_n/|\Delta|$
and the self-consistency equation coincides with the `gap equation'
for a bulk BCS superconductor. Keeping this analogy, convergence of
the Matsubara summation demands the inclusion of an energy cut-off
$E_c$ associated with the contact Coulomb interaction, analogous to the
Debye frequency scale in the superconductor, which sets the overall
scale of the order parameter. In the present case, the momentum
exchange associated with the Coulomb interaction is limited by the
screening length $\kappa$. As a result, the contact interaction, and
therefore, the Matsubara summation, must be cut-off at the energy
scale
\begin{equation}
  E_c = \Frac{p_F \kappa}{m} = \Frac{p_F}{m a_0}\; .
\label{eq:omega}
\end{equation}
In the high excitation limit where the photon coupling dominates, the
cutoff should be substituted with $\varepsilon_F$, the overall
bandwidth for electron-hole pairs.

At zero temperature, an explicit solution of the gap equation reveals
the gradual suppression of the condensate as a function of the
disorder strength $1/\tau_c |\Delta_0|$, leading to a complete
suppression of the order parameter when $2/\tau_c= |\Delta_0|$, where
$|\Delta_0|$ denotes the order parameter in the unperturbed system
(see Fig.~\ref{fig:engap}). (For a general review in the context of
the symmetry broken superconductors, see, e.g., Ref.~\cite{maki}.) As
with the disordered superconductor, the quasi-particle energy gap goes
to zero faster than the order parameter, exposing a gapless phase when
$2/\tau_c > e^{-\pi/4} |\Delta_0|$ or equivalently when the
dimensionless parameter $\zeta\equiv 1/\tau_c |\Delta|$ exceeds
unity. To understand how the phase behaviour of the Abrikosov-Gor'kov
equations translate to the polariton system, it is necessary to
explore the dependence of the chemical potential on the number of
excitations.

\subsection{Chemical Potential and Number of Excitations}
\label{sec:chemi}
Hitherto, we have absorbed the chemical potential into the definition
of an effective Fermi energy, $\varepsilon_F = (\mu - E_g)/2$,
assuming this to be the largest energy scale in the problem. However,
the chemical potential $\mu$ is itself fixed by the total number of
excitations in the system~\eqref{eq:nexci}. In the following, we will
use this condition to explore the parameter range of validity of the
approximations used to construct the theory, and to investigate the
phase diagram, treating $N_{\text{ex}}$ and the symmetry breaking
disorder potential $1/\tau_c$ as independent variables.

The validity of the model, and therefore of the self-consistency
equation~\eqref{eq:self2}, is restricted by the hierarchy of energy
scales contained in Eq.~\eqref{eq:scale}. Moreover, as implied by
Eq.~\eqref{eq:omega} above, the energy cut-off of the Coulomb
interaction is, itself, dependent on the chemical potential according
to the relation $E_c=\sqrt{\Ry (\mu-E_g)}$, where $\Ry$ is the Rydberg
energy of the electron-hole system. Taken together, the validity of
the mean-field theory requires $\varepsilon_F = (\mu - E_g)/2 > E_c$
(i.e. $(\mu - E_g)/\Ry > 4$), while to stay in the weak coupling
limit, we must have $E_c \gg |\Delta|$. In this limit, at zero
temperature, the total number of excitations $N_{\text{ex}}$, which is
proportional to the partial derivative of the free energy respect to
the chemical potential, can be expressed in terms of the chemical
potential as it follows:
\begin{equation*}
  N_{\text{ex}} = \Frac{g^2}{[g_c (\omega_c - \mu) + g^2]^2}
  |\Delta|^2 + \nu L^2\Frac{\mu - E_g}{2} \; .
\end{equation*}
The first term describes the total number of photons
$\mathcal{N}_{\text{ph}} = |\psi|^2$, while the second represents the
total number of particle-hole excitations in the system,
$\mathcal{N}_{\text{el}} = \nu L^2 \varepsilon_F$.  Here we have
neglected terms of order $O(|\Delta|/\varepsilon_F)$ associated with
the rearrangement of states close to the Fermi energy due to the
formation of the condensate. Taking into account the non-algebraic
dependence of the order parameter as well as the coupling constant
$g_c$ on $\mu$, the variation of the chemical potential with
$N_{\text{ex}}$ can be inferred only numerically.

\begin{figure}
\begin{center}
\includegraphics[width=1\linewidth,angle=0]{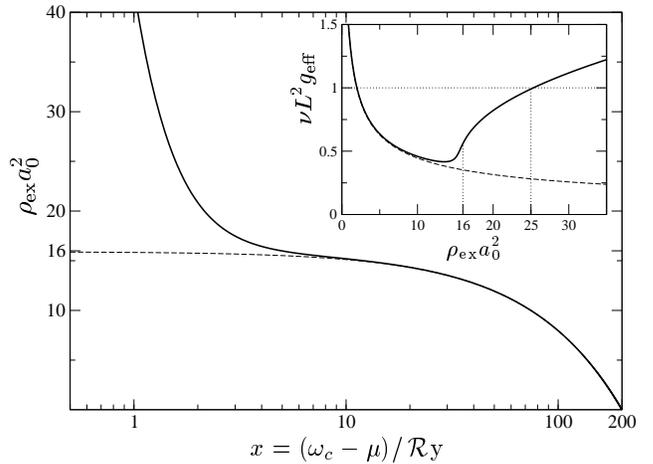}
\end{center}
\caption{\small Normalised density of excitations $\rho_{\text{ex}}
        a_0^2$ as a function of the rescaled chemical potential $x =
        (\omega_c - \mu)/\Ry$ (in logarithmic scale) for $d =
        (\omega_c - E_g)/\Ry = 200$, $\tilde{g} = 1$, $\lambda_c = 10$
        (bold line) and for $\tilde{g} = 0$ (dashed line). The value
        of the density at $x=0$ for $\tilde{g} = 0$, $ d/4\pi \simeq
        16$, has been marked. In this limit, all excitations appear as
        electrons and holes, allowing the population of states with
        energies in excess of $\omega_c$. The dependence from of the
        dimensionless effective coupling constant $\nu L^2
        g_{\text{eff}}$ on $\rho_{\text{ex}} a_0^2$ in both cases is
        shown in the inset. Note that the weak coupling limit imposes
        the constraint $\rho_{\text{ex}} a_0^2 < 25$.}
\label{fig:chemi}
\end{figure}

For simplicity, let us first consider the case of a clean quantum
well, $1/\tau_c = 0$. In this case, a solution of the zero temperature
self-consistency equation in the limit $\varepsilon_F > E_c \gg
|\Delta_0|$ obtains the BCS solution
\begin{equation*}
  |\Delta_0| = \Frac{E_c}{\sinh \left(1/\nu L^2
   g_{\text{eff}}\right)} \; .
\end{equation*}
Staying within the weak coupling regime, $\nu L^2 g_{\text{eff}} < 1$,
to check that the system remains in the high density phase, let us
recall that the Coulomb coupling constant is a function of the
chemical potential (see Eq.~\eqref{eq:ccons}). It is convenient to
introduce the dimensionless coupling constants~\eqref{eq:resca}. With
these definitions, the total density of excitations $\rho_{\text{ex}}
a_0^2 = N_{\text{ex}}/L^2 a_0^2$, takes the form
\begin{multline}
  \rho_{\text{ex}} a_0^2 \equiv \rho_{\text{el}} a_0^2 +
  \rho_{\text{ph}} a_0^2 = \Frac{d-x}{2\pi} \left\{\Frac{1}{2} +
  \right.\\ 
  \left. + \Frac{\tilde{g}^2}{(\tilde{g}^2 +
  x\tilde{g}_c(x))^2} \left[\sinh\left(\Frac{x}{\tilde{g}^2 +
  x\tilde{g}_c(x)}\right)\right]^{-2}\right\} \; ,
\label{eq:rscle}
\end{multline}
where $\rho_{\text{ph}} = \mathcal{N}_{\text{ph}}/L^2$ denotes the
photon density and $\rho_{\text{el}} =
\mathcal{N}_{\text{el}}/L^2\equiv 1/r_s^2$ denotes the \emph{total}
electron-hole density (and from which one can infer the $r_s$ value of
the electron-hole system). Inverted, Eq.~\eqref{eq:rscle} describes
the variation of the chemical potential (through $x$) as a function of
the material parameters $\tilde{g}$, $\lambda_c$, $d$ and the exciton
density $\rho_{\text{ex}}$.

In figure~\ref{fig:chemi}, the (normalised) density of excitations
$\rho_{\text{ex}} a_0^2$ is plotted against the renormalised chemical
potential $x$. When the density is low, the vast majority of
excitations are invested in particle-hole excitations and the chemical
potential scales linearly with the density, reflecting the constant
density of states in the quasi two-dimensional system. As the density
grows, the chemical potential $\mu$ converges on the photon band edge
$\omega_c$, i.e. the variable $x$ diminishes to zero.  Here, close to
the resonance, the excitations become increasingly photon-like and
$\rho_{\text{ex}} a_0^2$ diverges exponentially. The maximum density
of electron-hole excitations which can be added at $\mu = \omega_c$ in
the absence of the photon interaction, $\tilde{g} = 0$, is given by
$d/4\pi$. This value plays an important role, since, it represents the
value of the density at which the condensate becomes dominated by
photons. In fact, although both photons and electrons contribute to
the formation of the condensate (i.e. to $|\Delta_0|$), the electronic
contribution is dominant at low densities, while the screening of the
Coulomb interaction results in the growth of the photon contribution
at large densities. Such behaviour is illustrated most clearly in the
inset of Fig.~\ref{fig:chemi} where the effective coupling constant
$\nu L^2 g_{\text{eff}}$ is plotted against $\rho_{\text{ex}}
a_0^2$. Increasing the density of excitations above $d/4\pi$, the
effective coupling constant abruptly starts to increase until one
eventually reaches the strong coupling regime where the validity of
the microscopic Hamiltonian becomes uncertain. These characteristics
are reflected in figure~\ref{fig:nehph}, where the fraction of
\emph{total} electrons $\rho_{\text{el}}/ \rho_{\text{ex}}$ and
photons $\rho_{\text{ph}}/ \rho_{\text{ex}}$ are plotted as a function
of the density of excitations. Moreover, making use of the
equality~\eqref{eq:const}, the inset shows the variation of the
fraction of photons to excitons \emph{in the condensate},
\begin{equation}
  \Frac{\sqrt{\rho_{\text{ph}}a_0^2}}{N_{|\Sigma|}} =
  \Frac{\tilde{g}}{\tilde{g}_c} \Frac{\sqrt{4 \pi}}{x}\; ,
\label{eq:intoc}
\end{equation}
where $N_{|\Sigma|} = |\Sigma| \nu a_0^2$. Note that this ratio starts
to deviate from zero as soon as particles are added to the system. In
other words, even if small, there is a contribution from the photons
to the condensate at `low densities', when $\rho_{\text{ex}} a_0^2 \ll
d/4\pi$. Moreover, as already noted in the introduction, the increase
of this quantity with the increase of the total density of excitations
is due, on one hand, to the increase of the photon content as one
approaches the resonance while, on the other hand, is related to the
gradual decrease of Coulomb interaction and increase of the
exciton-photon interaction.

So far we have focused on the character of the polariton
condensate. In the next section, staying at the mean-field level, we
will explore its integrity when effected by the symmetry-breaking
disorder potential. Note that we have chosen particular values for the
parameters $d$, $\lambda_c$ and $\tilde{g}$ in such a way as to expose
a weak coupling regime in the high density limit. Changing their
values, the main characteristics of the phase diagram remain valid,
but the specific numbers can be varied. In particular, changing $d$,
the regions where electron-like (lower density) or photon-like (higher
density) excitations are dominant can be enlarged or reduced in
size. The range of the condensed region for $\rho_{\text{ex}} a_0^2 <
d/4\pi$ can be varied through $\lambda_c$, while changes in
$\tilde{g}$ characterise, when $\rho_{\text{ex}} a_0^2 > d/4\pi$,
variations in the gapless region and changes in the value at which the
strong coupling regime is entered.

\begin{figure}
\begin{center}
\includegraphics[width=1\linewidth,angle=0]{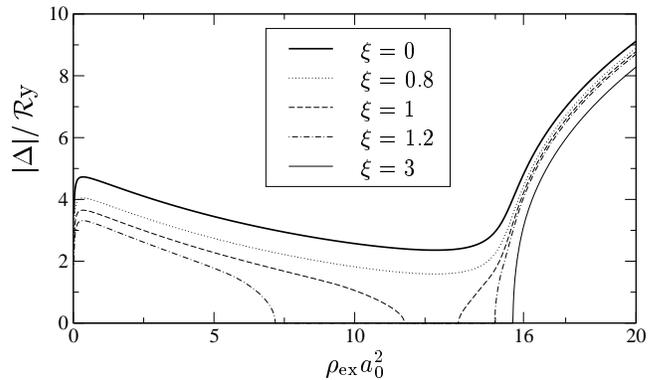}
\end{center}
\caption{\small Order parameter $|\Delta|/\Ry$ versus
        $\rho_{\text{ex}} a_0^2S$ for $\tilde{g} = 1$, $\lambda_c =
        10$ and $d = 200$ and different values of the disorder
        strength $\xi = 1/\tau_c \Ry$.}
\label{fig:dorpa}
\end{figure}
%
\subsection{Phase Diagram}
\label{sec:phase}
In the symmetry broken disordered system, the photon density (as
recorded in the magnitude of the order parameter) has to be rescaled
by the weight $|\Delta|/|\Delta_0|$. Fig.~\ref{fig:engap} shows the
variation of the order parameter against the disorder strength
$1/\tau_c |\Delta_0|$. Taking into account the dependence of the
unperturbed order parameter $|\Delta_0|$ on the chemical potential
(and therefore on the number of excitations), $|\Delta|/\Ry$ is
plotted in figure~\ref{fig:dorpa} against $\rho_{\text{ex}} a_0^2$ for
different values of $\xi =1/\tau_c \Ry$.

Let us consider first the dependence of the order parameter on the
excitation density in the clean case: Starting from $\rho_{\text{ex}}
a_0^2 \gtrsim 1$, $|\Delta_0|$ first slightly decreases, due to the
switching off of the Coulomb interaction, and then starts to increase
rapidly as
\begin{equation}
  |\Delta_0|/\Ry \simeq \sqrt{2\pi} \tilde{g} \sqrt{\rho_{\text{ex}}
   a_0^2} \; ,
\label{eq:ophig}
\end{equation}
for $\rho_{\text{ex}} a_0^2 > d/4\pi$. At these high values of the
density, the excitations in the condensate are mainly
photon-like. Therefore, increasing the disorder strength, it is not
surprising that the profile of the order parameter changes little. In
can be  shown analytically that, in this limit, the condition $1/\tau_c
|\Delta_0| > 1/2$ cannot be reached. At lower densities, however,
where the excitations are mainly electron-hole like, the order
parameter starts to close until, increasing the value of disorder, it
eventually vanishes everywhere. This behaviour is more evident from
the shape of the phase diagram (see Fig.~\ref{fig:phase}): In the
region of density where the excitations are mainly electron-hole like,
the condensed phase gets reduced, while, when the photon content
starts to dominate, the condensed region expands until eventually,
when $\rho_{\text{ex}} a_0^2 \simeq d/4\pi$, the disorder becomes
ineffective, and the line separating the two phases becomes
asymptotically straight.

Analogously, one can evaluate the density and disorder dependence of
the energy gap
\begin{equation*}
  \epsilon_{\text{gap}} = |\Delta| \left(1 - \zeta^{2/3}\right)^{3/2}
\end{equation*}
and determine the region where the excitation spectrum becomes
gapless, $\epsilon_{\text{gap}} = 0$ (which can be equivalently
established from the condition $\zeta = 1/\tau_c |\Delta| \ge 1$). As
a result, in the region dominated by the electronic excitations the
gapless phase is very small and lies just below the non-condensed
phase, while, when $\rho_{\text{ex}} a_0^2 > d/4\pi$, the line which
separates the condensed gapless region from the gapped one grows
approximatively as $\sqrt{2\pi} \tilde{g} \sqrt{\rho}$.

\begin{figure}
\begin{center}
\includegraphics[width=1\linewidth,angle=0]{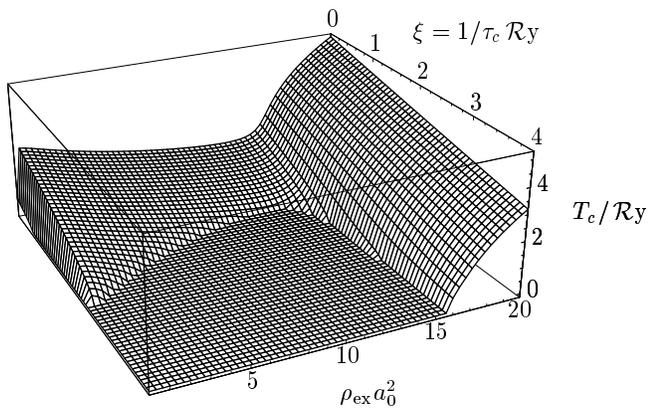}
\end{center}
\caption{\small Three-dimensional plot of the critical temperature,
        $T_c/\Ry$, versus the disorder strength, $\xi = 1/\tau_c \Ry$,
        and the excitation density, $\rho_{\text{ex}} a_0^2$
        ($\tilde{g} = 1$, $\lambda_c = 10$ and $d = 200$). The contour
        line where the critical temperature goes to zero separates the
        zero-temperature condensed phase from the non-condensed one
        and it coincides with the contour given in
        figure~\ref{fig:phase}.}
\label{fig:3dtri}
\end{figure}
%
\subsection{Finite Temperature}
\label{sec:finit}
Although a rigorous treatment of the finite temperature mean-field
theory would require an analysis of the thermal population of the
photon degrees of freedom, one can gain a qualitative insight by
approximating the chemical potential by its $T=0$ value. In this case,
the variation of the critical temperature $T_c$ with disorder can be
inferred from the gap equation through the
relation~\cite{abrikosov_gorkov,maki}
\begin{equation*}
  \ln\left(\Frac{T_c}{T_{c0}}\right) = \psi\left(\Frac{1}{2}\right) -
  \psi\left(\Frac{1}{2} + \Frac{1}{2\pi \tau_c T_c}\right) \; ,
\end{equation*}
where $T_{c0} = |\Delta_0| e^\gamma/\pi$, with $\gamma \simeq 0.577$,
is the critical temperature of the clean system, and $\psi (z) =
\Gamma'(z)/\Gamma(z)$ is the di-gamma function. Again, considering
explicitly the dependence of the bare order parameter $|\Delta_0|$ on
the density of excitations, it is possible to plot $T_c/\Ry$ as a
function of both the density $\rho_{\text{ex}} a_0^2$ and disorder
$\xi = 1/\tau_c \Ry$ (see Fig.~\ref{fig:3dtri}). The dependence of the
critical temperature on the density at given values of the disorder
closely mimics that of the order parameter $|\Delta|/\Ry$ given in
Fig.~\ref{fig:dorpa}: Starting from values of the densities
$\rho_{\text{ex}} a_0^2 \gtrsim 1$ the critical temperature starts to
decrease as the Coulomb interaction weakens. When the density is high
enough that the photon-like excitations are dominant, the critical
temperature increases again and becomes insensitive to disorder.

This concludes our discussion of the mean-field content of the
theory. In the following section, in order to explore the spectrum of
excitations above the condensate, we will develop an expansion of the
low-energy action in fluctuations around the mean-field.

\section{Collective Excitations}
\label{sec:excit}
In order to address important physical characteristics of the
condensed polariton system we need to go beyond the mean field
approximation and discuss excitation spectra. The excitations are
visible through optical absorption and photoluminescence, which can
provide important evidence for condensation in real systems. The phase
stiffness of the condensate is another important parameter, which
determines the transition temperature in the dilute limit - and in
two-dimensional systems at all densities. In 2D, since there is no
long range order (though there will be a
Berezinskii-Kosterlitz-Thouless transition), the velocity of the
Bogoliubov collective mode will determine the characteristic
temperature scale at which a finite droplet of condensate can emit
coherent light~\cite{keeling}.

Thus in this section we will explore the impact of fluctuations and
how they translate to collective phenomena. The latter can be divided
into separate contributions. In the first place, the system is
susceptible to both spatial and temporal fluctuations of the order
parameters, $|\Sigma|$ and $|\psi|$. In addition, there exist
fluctuations of the matrix fields $Q(\vect{r})$ which encode the
influence of mechanisms of quantum interference on the excitonic
degrees of freedom, the electrons and holes. In the following, we will
consider an approximation engaging the rearrangement of the
quasi-particle degrees of freedom in the condensate due to
fluctuations of the order parameter at the saddle-point level. The
impact of mesoscopic fluctuations and weak localisation phenomena on
the quasi-particle excitations is left as a subject for further
investigation. Drawing on the analogy with the superconducting system,
we note that related calculations of the pair susceptibility have been
performed in Ref.~\cite{eckern_pelzer} in the context of the
disordered superconducting film and, more directly, in
Ref.~\cite{smith_ambegaokar} in the context of a superconductor with
magnetic impurities.

In section~\ref{sec:model} it was noted that the single particle
Hamiltonian~\eqref{eq:singl} can be expressed in terms of the single
complex composite order parameter $\Delta = \Sigma + g\psi$. As such,
in considering fluctuations of the total action, it is convenient to
introduce a second independent field $\Gamma =\Sigma - g\psi$ and
express the fluctuations of the excitonic and photonic order
parameters in terms of the fluctuations of $\Delta$ and $\Gamma$, viz.
\begin{equation}
\begin{split}
  \Delta (\vect{r}, \tau) &= |\Delta| + \delta \Delta^{\smc{l}}
  (\vect{r}, \tau) + i \delta \Delta^{\smc{t}} (\vect{r}, \tau)\\
  \Gamma (\vect{r}, \tau) &= |\Gamma| + \delta \Gamma^{\smc{l}}
  (\vect{r}, \tau) + i \delta \Gamma^{\smc{t}} (\vect{r}, \tau)\; ,
\end{split}
\label{eq:fluct}
\end{equation}
where $|\Delta|$ and $|\Gamma|$ denote the mean-field expectation
values.  Remembering that, at the mean-field level, the phases of
$\Sigma$ and $\psi$ are locked (see Eq.~\eqref{eq:selfo}), one can set
$|\Delta| = |\Sigma| + g|\psi|$ and $|\Gamma| = |\Sigma| - g|\psi|$,
while $|\Sigma|$ and $|\psi|$ are constrained by the
relation~\eqref{eq:const}. In Eq.~\eqref{eq:fluct}, longitudinal
(\smc{l}) and transverse (\smc{t}) fluctuations represent respectively
fluctuations of the amplitude and of the phase. Since our model is
invariant under a global gauge transformation of the
phase~\eqref{eq:gauge}, we can anticipate that the spectrum of
fluctuations will exhibit one massless Goldstone mode and three
massive ones. To develop a theory for fluctuations of the condensed
field $\Delta$, two of the three massive fluctuations associated with
the field $\Gamma$ will be integrated out.

Fluctuations of the order parameters $\Delta$ and $\Gamma$ are not
divorced from fluctuations of the fields $Q$. Indeed, through the
coupling, one can expect the fluctuations of the order parameter to
acquire a diffusive nature in the disordered system. Moreover, in the
gapless phase, the existence of magnetic impurities should induce a
mechanism of dissipation.  To capture these effects, one must
determine the rearrangement of the field (or self-energy) $Q$ due to
fluctuations of the order parameter.

\begin{figure}
\begin{center}
\includegraphics[width=1\linewidth,angle=0]{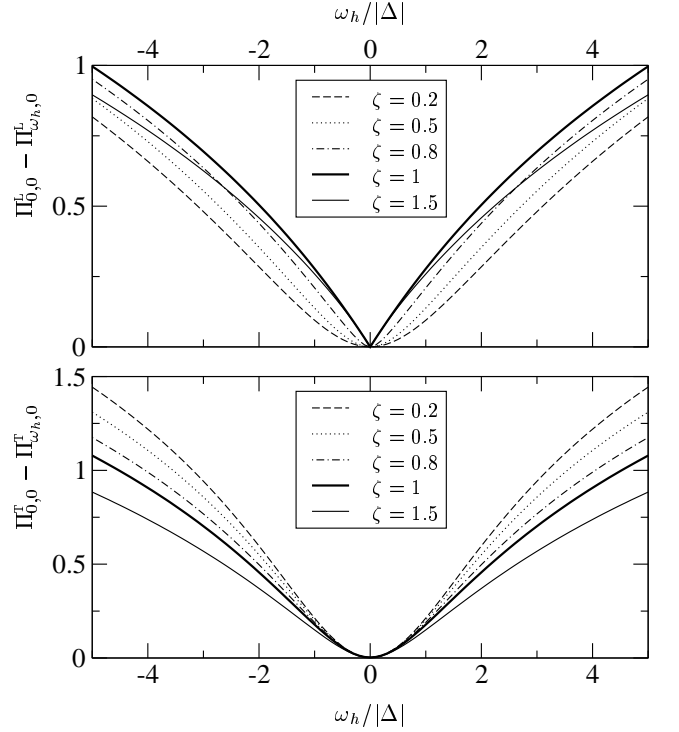}
\end{center}
\caption{\small Zero temperature plot of the zero momentum
        longitudinal $\Pi_{0,0}^{\smc{l}} -
        \Pi_{\omega_h,0}^{\smc{l}}$ and transverse
        $\Pi_{0,0}^{\smc{t}} - \Pi_{\omega_h,0}^{\smc{t}}$ components
        versus the rescaled frequency $\omega_h/|\Delta|$, for
        different values of the disorder strength $\zeta = 1/\tau_c
        |\Delta|$.}
\label{fig:pilto}
\end{figure}

Denoting the homogeneous saddle-point solution associated with the
mean-field as $\bar{Q}$, let us represent the rearrangement due to
fluctuations as
\begin{equation}
  Q (\vect{r}) = e^{W(\vect{r})/2} \bar{Q} e^{-W(\vect{r})/2} \; ,
\label{eq:param}
\end{equation}
where the generators obey the symmetry relation $\{\bar{Q} , W
(\vect{r}\} = 0$. Although the calculation of the rotation $W$ in
linear response is straightforward, the technical details are involved
and not illuminating. As such, they have been included as an
appendix. Here, we note only that, at the level of the linear
response, the longitudinal and transverse components of the order
parameter fluctuations do not get mixed by the $Q$-matrix action. By
contrast, expanding $S_{\text{ph}} + S_{\Sigma}$ up to second order in
$\delta \Delta^{\smc{l,t}}$ and $\delta \Gamma^{\smc{l,t}}$, the
(massive) fluctuations of the latter can be integrated out revealing a
contribution which mixes the longitudinal and transverse
response. Gathering all of the contributions together, the total
action takes the form
\begin{equation*}
  S_{\text{tot}} = S^{\smc{ag}} [|\Delta| , g_{\text{eff}}] + \delta
  S_{\text{tot}}\; ,
\end{equation*}
where $S^{\smc{ag}} [|\Delta| , g_{\text{eff}}]$ represents the
mean-field Abrikosov and Gor'kov action for a superconductor with
magnetic impurities~\eqref{eq:mfabg} and $\delta S_{\text{tot}}$ has
the quadratic form
\begin{equation}
  \Frac{\delta S_{\text{tot}}}{\beta \nu L^2} = \sum_{\omega_h,\vect{p}}
  \begin{pmatrix}
    \delta\Delta^{\smc{l}} \\
    \delta\Delta^{\smc{t}}
  \end{pmatrix}^{\mathsf{T}}_{-\omega_h , -\vect{p}}
  G^{-1}_{\omega_h , \vect{p}}
  \begin{pmatrix}
    \delta\Delta^{\smc{l}}\\
    \delta\Delta^{\smc{t}}
  \end{pmatrix}_{\omega_h , \vect{p}}\; .
\label{eq:secon}
\end{equation}
Here the kernel $G^{-1}_{\omega_h , \vect{p}}$ denotes the inverse of
the thermal Green's function or susceptibility:
\begin{equation}
  G^{-1}_{\omega_h , \vect{p}} =
  \begin{pmatrix}
     - \Pi^{\smc{l}}_{\omega_h , \vect{p}} + \Theta_{\omega_h ,
    \vect{p}} & \Lambda_{-\omega_h , \vect{p}}\\ 
     \Lambda_{\omega_h , \vect{p}} & - \Pi^{\smc{t}}_{\omega_h ,
    \vect{p}} + \Theta_{\omega_h , \vect{p}}
  \end{pmatrix} \; .
\label{eq:kerne}
\end{equation}
We refer to appendix~\ref{sec:secon} for details and the explicit
expressions of the Green's function elements. Let us note here that,
if we restrict attention to the condensate phase $|\Delta| \ne 0$, the
self-consistency equation~\eqref{eq:self2} translates to the condition
\begin{equation}
  \Theta_{0,0} \equiv \Frac{1}{\nu L^2 g_{\text{eff}}} =
  \Pi^{\smc{t}}_{0,0} \; ,
\label{eq:costr}
\end{equation}
rendering the transverse fluctuations of the quasi-particle degrees of
freedom massless. By contrast, when $|\Delta| = 0$, the longitudinal
and transverse kernels coincide, $\Pi^{\smc{l}}_{\omega_h,\vect{p}} =
\Pi^{\smc{t}}_{\omega_h,\vect{p}}$ and acquire a mass, which, at zero
temperature, is given by
\begin{equation*}
  \Theta_{0,0} - \Pi^{\smc{l,t}}_{0,0} = \ln\left(\Frac{2}{|\Delta_0|
  \tau_c}\right)\; .
\end{equation*}
As expected from the mean-field analysis, when $2/|\Delta_0| \tau_c =
1$ the mass disappears and the system undergoes a transition to a
condensate phase.

The form in Eq.~\eqref{eq:secon} is positive definite, determining the
stability of the mean-field solution against fluctuations. Moreover it
is already clear from the form of the kernel~\eqref{eq:kerne} and from
the constraint~\eqref{eq:costr} that the excitation spectra of the
system admits a massless Goldstone mode and a massive one. In fact,
considering the zero frequency and zero momentum component, the
anti-diagonal term $\Lambda_{0 , 0}$ vanishes together with the
transverse one, $(G^{-1}_{0 , 0})_{22}$, while the longitudinal
component $(G^{-1}_{0 , 0})_{11}$ remains finite.

In fact, the precise functional forms
$\Pi^{\smc{l,t}}_{\omega_h,\vect{p}}$ coincide with those found for a
disordered superconductor in the presence of magnetic
impurities~\cite{smith_ambegaokar}. Their frequency dependence is
plotted in figure~\ref{fig:pilto}: With $\Pi_{\omega_h,0}^{\smc{t}}$,
no qualitative change is associated with the transition to the gapless
region for $\zeta \ge 1$; the quadratic character of the dispersion at
small frequencies is preserved. By contrast, as the system enters the
gapless phase $\zeta\ge 1$, with the longitudinal component
$\Pi_{\omega_h,0}^{\smc{l}}$ exhibits a dissipative dependence scaling
linearly with frequency.

In order to elucidate this behaviour further, it is convenient to
affect a gradient expansion of the kernel~\eqref{eq:kerne}, whereupon,
at zero temperature, one finds
%
\begin{widetext}
%
\begin{equation}
  G^{-1}_{\omega_h , \vect{p}} \simeq
  \begin{pmatrix}
     c_1 + \Frac{1}{2} a_1 \Frac{D \vect{p}^2}{|\Delta|} +
    \Frac{1}{2\zeta} \Frac{|\omega_h|}{|\Delta|} \theta (\zeta - 1)+
    \Frac{1}{2} b_1 \left(\Frac{\omega_h}{|\Delta|}\right)^2 & - f
    \Frac{\omega_h}{|\Delta|}\\
    f \Frac{\omega_h}{|\Delta|} & \Frac{1}{2} a_2 \Frac{D
    \vect{p}^2}{|\Delta|} + \Frac{1}{2} b_2
    \left(\Frac{\omega_h}{|\Delta|}\right)^2 
  \end{pmatrix} \; ,
\label{eq:expan}
\end{equation}
\end{widetext}
%
with
\begin{align}
\nonumber
  a_1 &= a_{\smc{l}} (\zeta) + a & a_2 &= a_{\smc{t}} (\zeta) + a\\
  b_1 &= b_{\smc{l}} (\zeta) + b & b_2 &= \frac{1}{2} + b \; .
\label{eq:coeff}
\end{align}
The coefficients $c_1 = c_{\smc{l}} (\zeta)$, $a_{\smc{l,t}} (\zeta)$
and $b_{\smc{l,t}} (\zeta)$ derive from the expansion of the kernels
$\Pi^{\smc{l,t}}_{\omega_h,\vect{p}}$~\cite{moir}. Although the
explicit expression for their dependency on the disorder strength
$\zeta$ can be rigorously derived (or equivalently can be numerically
inferred from figure~\eqref{fig:pilto}), we note here that, as soon as
$|\Delta| \ne 0$, these parameters are of $O (1)$ and in particular,
at zero disorder strength $\zeta = 0$, they are given by $c_{\smc{l}}
(0) = 1$, $a_{\smc{l}} (0) =\pi/4$, $a_{\smc{t}} (0)= \pi/2$,
$b_{\smc{l}} (0) = 1/6$.  More significantly, in the gapless region
$\zeta > 1$, the longitudinal kernel acquires a dissipative term and
the coefficient $b_{\smc{l}} (\zeta)$ changes of sign and becomes
negative. In contrast, making use of the dimensionless parameters
introduced in Eq.~\eqref{eq:resca}, the expansion of the kernels
$\Theta_{\omega_h , \vect{p}}$ and $\Lambda_{\omega_h , \vect{p}}$,
gives

\begin{align*}
  a &= \Frac{\tilde{g}^2}{[\tilde{g}^2 + \tilde{g}_c (x) x]^2}
  \Frac{c^2}{D \omega_c} \left(\Frac{|\Delta|}{\Ry}\right)\\
  b &= \Frac{2\tilde{g}^2\tilde{g}_c (x)}{[\tilde{g}^2 + \tilde{g}_c (x)
  x]^3} \left(\Frac{|\Delta|}{\Ry}\right)^2\\ 
  f &= \Frac{\tilde{g}^2}{[\tilde{g}^2 + \tilde{g}_c(x) x]^2}
  \left(\Frac{|\Delta|}{\Ry}\right)\; .
\end{align*}
%
\subsection{Gapped Phase}
\label{sec:gappe}
\begin{figure}
\begin{center}
\includegraphics[width=1\linewidth,angle=0]{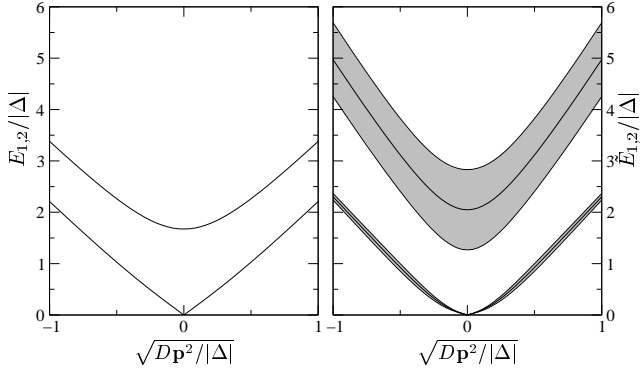}
\end{center}
\caption{\small Dispersion of the collective excitation spectrum in
        the gapped $\zeta < 1$ (left) and gapless $\zeta > 1$ (right)
        regions. Here, we have chosen the particular values $c_{1} =
        0.8$, $a_1=7.1$, $a_2=7.7$, $b_1=1$, $b_2=1.2$ and $f=0.6$
        (corresponding to $\rho_{\text{ex}} a_0^2 = 16$, $\xi =
        1/\tau_c\Ry = 1$ and $D\omega_c/c^2 = 0.1$) in the first case,
        and $c_{1} = 0.1$, $a_1=7.6$, $a_2=7.6$, $b_1=0.4$, $b_2=1.1$
        and $f=0.7$ (corresponding to $\rho_{\text{ex}} a_0^2 = 16$,
        $\xi = 4$ and $D\omega_c/c^2 = 0.1$) in the second (note that
        this choice masks the actual linear dispersion of the massless
        mode). When $\zeta > 1$, both the massive and massless
        excitations acquire a finite life time with, respectively the
        first finite and almost constant, while the second exhibits a
        quadratic dispersion and therefore null at zero
        momentum. Their line widths, $\Re \tilde{E}_{1,2} \pm \Im
        \tilde{E}_{1,2}$, are explicitly plotted.}
\label{fig:disco}
\end{figure}

The spectrum of the collective excitations is determined by the poles
of the matrix $G_{\omega_h , \vect{p}}$ and therefore by the zeros of
the expression $\det (G^{-1}_{\omega_h , \vect{p}})$. Let us consider
first the gapped phase, $\zeta < 1$, where the determinant of the
kernel can be evaluated exactly:
\begin{equation*}
  \det \left(G^{-1}_{\omega_h , \vect{p}}\right) =
  \left(\Frac{\omega_h^2}{|\Delta|^2} +
  \Frac{E_{1}^2}{|\Delta|^2}\right)\left(\Frac{\omega_h^2}{|\Delta|^2}
  + \Frac{E_{2}^2}{|\Delta|^2}\right) \; .
\end{equation*}
The energies $E_{1,2}$, which can be shown for each value of the
momentum to be real, are plotted in figure~\ref{fig:disco}. From this
we can infer that the spectrum of excitations, which is obtained by
analytically continuing to real energies (i.e., substituting $\omega_h
\mapsto i\varepsilon$), is given by a linear (massless) branch and a
quadratic (massive) branch, where, expanding for small values of
momentum, the gap of the massive mode and the `velocity' associated
with the massless mode are respectively given by:
\begin{align*}
  \Frac{E_1}{|\Delta|} &\Simiq_{\vect{p} \to 0} \sqrt{\Frac{2b_2 c_1 +
  4f^2}{b_1 b_2}} \\
  \Frac{E_2}{|\Delta|} &\Simiq_{\vect{p} \to 0} \sqrt{\Frac{a_2
  c_1}{b_2 c_1 + 2f^2}} \sqrt{\Frac{D \vect{p}^2}{|\Delta|}} \; .
\end{align*}

In order to gain some physical insight into these parameters, once
they are expressed in terms of the coefficients~\eqref{eq:coeff}, two
opposite limits can be identified. In the first case, when $1 <
\rho_{\text{ex}} a_0^2 \ll d/4\pi$ (i.e. for $x \gg 1$, $\tilde{g}_c
(x) x \gg \tilde{g}$ and $|\Delta|/\Ry < O(1)$) one expects
$a_{\smc{l,t}} \gg a$ and $b_{\smc{l,t}} \gg b,f$, whereupon
\begin{align*}
  E_1 &\Simiq_{\vect{p} \to 0} \sqrt{\Frac{2
  c_{\smc{l}}}{b_{\smc{l}}}} |\Delta| \To_{\zeta \to 0} 2 \sqrt{3}
  |\Delta_0|\\
  E_2 &\Simiq_{\vect{p} \to 0} \sqrt{\Frac{a_{\smc{t}}}{b_{\smc{t}}}}
  \sqrt{D |\Delta|} |\vect{p}| \To_{\zeta \to 0} \sqrt{\pi} \sqrt{D
  |\Delta_0|} |\vect{p}|\; .
\end{align*}
Therefore, in the gapped phase, for values of the density at which the
excitations are largely electronic in character (see
Fig.~\ref{fig:nehph}), the gap for the massive collective excitations
is proportional to the composite order parameter $|\Delta|$, while the
velocity of the massless excitations is given by the product of the
coherence length $\xi_{\text{coh}} = \sqrt{D/|\Delta|}$ and the order
parameter~\cite{comment}.

In the opposite limit, $\rho_{\text{ex}} a_0^2 \gg d/4\pi$ (i.e. for
$x \to 0$, and, taking into account the density dependence of the
order parameter in this limit~\eqref{eq:ophig}, for $d/4\pi >
\tilde{g}^2 /\tilde{g}_c (x)$), we can use the approximations $a \gg
a_{\smc{l,t}}$ and $b,f \gg b_{\smc{l,t}}$, which translates to
\begin{align*}
  E_1 &\Simiq_{\vect{p} \to 0} \Frac{\tilde{g}^2}{\tilde{g}_c (x)} \Ry
  \\ 
  E_2 &\Simiq_{\vect{p} \to 0} \tilde{g} \sqrt{\Frac{c_{\smc{l}}}{2
  (\omega_c/\Ry) (1 + c_{\smc{l}})}} c \vect{p}\; .
\end{align*}
When the density is so high that photons dominate, but still $g_c (x)
\ne 0$, the gap in the collective excitation spectrum is given by the
ratio of the square of the photon coupling to the Coulomb coupling
constant, while the velocity is proportional to the velocity of light
$c$. Note that in absence of Coulomb interaction, $g_c \equiv 0$, the
coefficient $b$ vanishes and the situation is different: the gap is in
fact proportional to $|\Delta| = g |\psi|$ in the first limit and to
$|\psi|^2/\nu L^2$ in the very high density regime.
%
\subsection{Gapless Phase}
\label{sec:gaple}
Let us consider now the gapless region $\zeta > 1$. In the leading
approximation, one may neglect the quadratic term $b_{1}
\omega_h^2/|\Delta|^2$, obtaining
\begin{gather*}
  \det \left(G^{-1}_{\omega_h , \vect{p}}\right) = \left(i
  \Frac{|\omega_h|}{|\Delta|} + \Frac{\tilde{E}_1}{|\Delta|}\right)
  \left(-i \Frac{|\omega_h|}{|\Delta|} -
  \Frac{\tilde{E}_2}{|\Delta|}\right)\\
  \Frac{\tilde{E}_{1,2}}{|\Delta|} \Simiq_{\vect{p} \to 0} \pm
  \sqrt{\Frac{a_2 c_1}{b_2 c_1 + 2f^2}} \sqrt{\Frac{D
  \vect{p}^2}{|\Delta|}} + i \Frac{a_2}{4 \zeta (b_2 c_1 + 2f^2)}
  \Frac{D \vect{p}^2}{|\Delta|}\; .
\end{gather*}
Therefore, analytically continuing to real energies ($|\omega_h| \to
i\varepsilon$), the massless modes $E_{2}$ acquires a positive
imaginary part with a quadratic dispersion in terms of the momentum
$\vect{p}$. In other words the massless mode has a finite life time
(or line width) whenever $\vect{p} \ne 0$. In order to evaluate the
life time of the massive modes $E_1$, one has to restore the quadratic
term $b_{1}\omega_h^2/|\Delta|$. Considering for simplicity a value of
the density $\rho_{\text{ex}} a_0^2$ and the disorder $\xi = 1/\tau_c
\Ry$ for which the coefficient $b_1$ is overall positive, the spectrum
is plotted in figure~\ref{fig:disco}. (Note that close to the
transition between the gapped and the gapless region, the coefficient
$b_1$ can become negative, in which case, to reestablish stability,
one has to introduce the next order in frequency of the expansion.)

We finally note that, in the non-condensed region $|\Delta| = 0$
(i.e., at zero temperature, for values of the disorder strength
$2/\tau_c |\Delta_0| > 1$), the kernel acquires a mass proportional to
$\ln(2/\tau_c |\Delta_0|)$, while the massless modes disappear.

\subsection{Number of Photonic Occupied Modes}
\label{sec:numbe}
Experimentally, the spectrum of collective excitations can be accessed
indirectly by measuring the angle resolved photoluminescence intensity
associated with the small leakage of photons from the cavity. The
latter provides a convenient way of mapping the occupation density of
the photon modes,
\begin{equation*}
  \mathcal{N}_{\text{ph}} (\vect{p}) = \lim_{\eta \to 0^+} \langle
  \psi^\dag_{\vect{p}} (\tau + \eta) \psi_{\vect{p}} (\tau)\rangle \;
  ,
\end{equation*}
where the in-plane quantum well momentum is determined by the angle
$\phi$ at which the light is detected, $|\vect{p}| = (\omega_c/c) \tan
\phi$. Taken at the level of the mean-field approximation, the
transition to the condensed phase would be associated with the
development of a sharp coherent peak in the density distribution at
$\vect{p}=0$. However, in the quasi two-dimensional quantum well
geometry, the proliferation of massless phase modes destroys true long
range order. Instead, the bulk transition to the condensed phase is of
Berezinskii-Kosterlitz-Thouless type~\cite{chaikin_lubensky}. In this
case, one expects the bulk transition to be reflected in the
development of power-law correlations in the density
distribution. Strictly speaking, to infer the characteristic power-law
dependence of the correlations, one should restore gauge invariance of
the action under global phase rotations. However, guided by a parallel
investigation of trapped excitons in Ref.~\cite{keeling}, one can
infer the nature of the long ranged correlations simply from the
spectrum of the phase mode determined above.

In the present case the calculation is complicated slightly by the
fact that, to establish the impact of the massless collective
fluctuations on the photon distribution, one must find their
projection onto the phase mode associated with the photon field. This,
in turn, introduces a `form factor' which reflects the superfluid
stiffness. Then, when parameterized as $\psi (\vect{r}) = |\psi
(\vect{r})|e^{i \theta (\vect{r})}$, discarding massive fluctuations,
one obtains the characteristic power-law
divergence~\cite{footnote_fluct}
\begin{equation}
  \langle \theta_{-\vect{p}} \theta_{\vect{p}} \rangle
  \Simiq_{\vect{p} \to 0} \Frac{k_{\smc{b}} T}{\rho_s (\vect{p}
  L)^2}\; ,
\label{eq:phpro}
\end{equation}
with the (superfluid) stiffness given by
\begin{equation*}
  \rho_s = \Frac{[\tilde{g}^2 + \tilde{g}_c (x) x]^2}{\tilde{g}^2}
  \Frac{a_2}{2} \left(\Frac{\xi_{\text{coh}}}{L}\right)^2 \Ry \psi^2
  \; ,
\end{equation*}
where $\xi_{\text{coh}} = \sqrt{D/|\Delta|}$ is the superconducting
coherence length. The corresponding spatial correlation function
assumes the characteristic logarithmic divergence at long distances:
\begin{equation*}
  g (\vect{r}) \equiv \Frac{1}{2} \langle\left[\theta (\vect{r}) -
  \theta (0)\right]^2\rangle \Simiq_{|\vect{r}| \to \infty}
  \Frac{k_{\smc{b}} T}{2\pi \rho_s} \ln
  \left(\Frac{|\vect{r}|}{\xi_{T}}\right)\; ,
\end{equation*}
where $\xi_{T} \propto \xi_{\text{coh}} |\Delta|/k_{\smc{b}} T$
represents the thermal length. Restoring gauge invariance, and noting
that the dominant contribution to the amplitude mode comes from
mean-field, we obtain:
\begin{equation*}
  \mathcal{N}_{\text{ph}} (\vect{p}) = |\psi|^2 \int
  \Frac{d\vect{r}}{L^2} e^{i \vect{p} \cdot \vect{r} - g(\vect{r})} \;
  .
\end{equation*}
This integral can be exactly evaluated in two-dimensions, giving, for
temperatures $\eta = k_{\smc{b}} T/2 \pi \rho_s < 2$, the result:
\begin{equation}
  \mathcal{N}_{\text{ph}} (\vect{p}) = 4 \pi \Frac{|\psi|^2}{L^2}
  \left(\Frac{\xi_T}{2}\right)^{\eta} \Frac{\Gamma (1 -
  \eta/2)}{\Gamma(\eta/2)} \Frac{1}{|\vect{p}|^{2 - \eta}} \; .
\end{equation}
As noted at the end of section~\ref{sec:gaple}, when $|\Delta| = 0$,
the amplitude and phase modes coincide and, in particular, the phase
mode acquires a mass proportional to the distance from the critical
point. Here, in the non-condensed phase, the density distribution
asymptotes to a constant at small $\vect{p}$.

\section{Conclusions}
\label{sec:concl}
In this paper, we have investigated the properties of a model for
cavity polaritons in the high density regime, where the interplay
between Coulomb interaction, the photon mediated coupling and the
effect of disorder in the quantum well gives rise to different domains
in the phase diagram. Changing the density, we have shown that the
character of the condensate changes from being excitonic in character
to photonic. The effective coupling constant is determined by the sum
of the direct screened Coulomb interaction and the effective
interaction induced by the exchange of photons. When the condensate
has largely excitonic character, we have shown that the `charged'
disorder presents a pair-breaking channel which destabilises the
condensate phase. Increasing the density of excitations, the system
moves towards a region dominated by photons. Here, the increasing
insensitivity of the condensate to disorder inhibits the complete
quenching of coherence, while leaving open the possibility of a
gapless region where the condensate manifests the properties of a
semiconductor laser. In the condensed phase, across the whole range of
densities, we have determined the spectrum of collective excitations.

The parameter regimes we have explored are in principle accessible in
semiconductor microcavities.  In III-V and II-VI semiconductors,
microcavities have been constructed with measured values of the Rabi
splitting approaching $30 \text{meV}$~\cite{lesidang}, and even larger
values for an organic semiconductor~\cite{lidzey}. Since the exciton
Rydberg and inverse density of states take values similar to or
smaller than this energy (depending on material), the dimensionless
photon coupling $\tilde{g}$ can be made of order unity. To reach the
high density `plasma' regime for the electron-hole system, one must
obtain electron-hole densities $\rho_{\text{el}} a_0^2$ of order unity
or greater, but this is indeed the typical operating density of a
semiconductor laser. Indeed semiconductor lasers are conventionally
modelled as a degenerate electron-hole plasma, though of course their
operation is conventionally at temperatures somewhat larger than the
Rydberg. Perhaps more important than their temperature is that real
lasers are open systems and thus have dynamical decoherence and
dephasing processes that we have not included here. To the extent that
pair-breaking by static disorder mimics dynamically induced
decoherence, we may interpret the pair-breaking parameter $\zeta$ as
the dephasing rate $T_2^{-1}$ introduced into laser rate
equations. With this identification parameters for a conventional
laser~\cite{sclaser} are indeed well in the limit $\zeta \gg 1$.

As a final remark, for the clean system, the model considered in this
work bears similarity with that discussed in the context of atomic
condensation. The polariton Hamiltonian considered in this work has
been used by many to model the atom/molecule interaction in a Fermi
system positioned close to a Feshbach resonance (see, e.g.,
Ref.~\cite{timmermans_holland_ohashi_griffin}). With an appropriate
dictionary of correspondence, both the mean-field content of the
theory as well as the spectrum of collective excitations can be
compared in the two systems.

\appendix

\section{Second Variation}
\label{sec:secon}
As outlined in section~\ref{sec:excit}, the second variation $\delta
S_{\text{tot}} = \delta S[Q] + \delta S_{\text{ph} + \Sigma}$ of the
total action is given by two contributions, one coming from the action
for the semiclassical Green function~\eqref{eq:actio} and the other
from the photonic~\eqref{eq:phact} and excitonic~\eqref{eq:eiact}
actions. Starting with the first contribution, we consider the
parameterization~\eqref{eq:param} for the $Q (\vect{r})$ matrix and
linearize the equation of motion~\eqref{eq:motio} in the fluctuation
terms $W (\vect{r})$ and $\delta \Delta^{\smc{l,t}}_{\omega_h}
(\vect{r})$,
\begin{multline*}
  -D \nabla^2 W + [\hat{\mathcal{O}}, \bar{Q}W] -
  \Frac{1}{2\tau_c} [\bar{Q} , \sigma_3 \sigma_3^{\smc{cc}} \bar{Q} W
  \sigma_3 \sigma_3^{\smc{cc}}] \\
  = - [\delta \Delta^{\smc{l}} \sigma_2 + \delta \Delta^{\smc{t}}
  \sigma_1, \bar{Q}]\; ,
\end{multline*}
where the operator 
\begin{equation*}
  \hat{\mathcal{O}} = \hat{\epsilon} \sigma_3 \sigma_3^\smc{cc} -
  |\Delta| \sigma_2 + \Frac{1}{2\tau_c} \sigma_3 \sigma_3^{\smc{cc}}
  \bar{Q} \sigma_3 \sigma_3^{\smc{cc}}\; ,
\end{equation*}
generates the mean-field equation of motion, $[\hat{\mathcal{O}} ,
\bar{Q}] = 0$. The equation for $W (\vect{r})$ can be easily solved in
term of the components $W = w_0 \openone + \vect{w} \cdot
\vectgr\sigma \sigma_3^{\smc{cc}}$. Then, this solution can be
substituted back into the action~\eqref{eq:actio}, which expanded up
to the second order, reads as $S[Q] = S[\bar{Q}] + \delta S[Q]$, where
\begin{equation*}
  \delta S[Q] = - \Frac{\pi \nu}{8} \int d\vect{r} \tr\left([\delta
  \Delta^{\smc{l}} \sigma_2 + \delta \Delta^{\smc{t}} \sigma_1,
  \bar{Q}] W\right)\; .
\end{equation*}
Substituting the solutions for $w_0$ and $\vect{w}$, one obtains
\begin{multline}
  \delta S[Q] = - \nu L^2 \beta \sum_{\omega_h , \vect{p}}
  \left[\delta \Delta_{-\omega_h , -\vect{p}}^{\smc{l}}
  \Pi^{\smc{l}}_{\omega_h , \vect{p}} \delta \Delta_{\omega_h ,
  \vect{p}}^{\smc{l}}\right. \\
  \left. + \delta \Delta_{-\omega_h ,
  -\vect{p}}^{\smc{t}} \Pi^{\smc{t}}_{\omega_h,\vect{p}}
  \delta \Delta_{\omega_h , \vect{p}}^{\smc{t}}\right] \; ,
\label{eq:1cont}
\end{multline}
where the kernels $\Pi^{\smc{l,t}}_{\omega_h,\vect{p}}$ are given by:
\begin{widetext}
\begin{equation}
\begin{split}
  \Pi^{\smc{l,t}}_{\omega_h,\vect{p}} &= \Frac{\pi}{\beta |\Delta|}
  \sum_{\epsilon_n} \Omega^{\smc{l,t}}_{\vect{p}} (u (\epsilon_n) , u
  (\epsilon_n + \omega_h))\\
  \Omega^{\smc{l},\smc{t}}_{\vect{p}} (u_1 , u_2) &= \Frac{1 +
   \Frac{u_1 u_2 \mp 1}{\sqrt{1 + u_1^2} \sqrt{1 +
   u_2^2}}}{\Frac{D\vect{p}^2}{|\Delta|} + \sqrt{1 + u_1^2} +
  \sqrt{1 + u_2^2} - \zeta \left(1 - \Frac{u_1 u_2 \mp 1}{\sqrt{1 +
  u_1^2} \sqrt{1 + u_2^2}}\right)} \; ,
\end{split}
\label{eq:piexp}
\end{equation}
\end{widetext}
%

Turning to the second contribution, one has to expand $S_{\text{ph}} +
S_{\Sigma}$ around the mean field value of the photon field $|\psi|$
and of excitonic order parameter $|\Sigma|$. As suggested in
section~\ref{sec:excit}, it is convenient to introduce a new order
parameter $\Gamma (\vect{r} , \tau)$ and parameterise the fluctuations
as in Eq.~\eqref{eq:fluct}. Once expanded up to the second order, the
variables $\delta\Gamma^{\smc{l},\smc{t}}_{\pm \omega_h}$ can be
eliminated via a Gaussian integral. In this way, $S_{\text{ph}} +
S_{\Sigma} = \beta (\omega_c - \mu) |\psi|^2 + \beta |\Sigma|^2/g_c +
\delta S_{\text{ph} + \Sigma}$, where
\begin{multline}
  \delta S_{\text{ph} + \Sigma} = \beta \sum_{\omega_h , \vect{p}}
  \left\{\delta\Delta^{\smc{l}}_{-\omega_h , -\vect{p}}
  \Theta_{\omega_h , \vect{p}} \delta\Delta^{\smc{l}}_{\omega_h ,
  \vect{p}} \right. \\ 
  \left. + \delta\Delta^{\smc{t}}_{-\omega_h , -\vect{p}}
  \Theta_{\omega_h , \vect{p}} \delta\Delta^{\smc{t}}_{\omega_h ,
  \vect{p}} + \delta\Delta^{\smc{l}}_{-\omega_h , -\vect{p}}
  \Lambda_{-\omega_h, \vect{p}} \delta\Delta^{\smc{t}}_{\omega_h ,
  \vect{p}}\right. \\
  \left. + \delta\Delta^{\smc{t}}_{-\omega_h , -\vect{p}}
  \Lambda_{\omega_h , \vect{p}} \delta\Delta^{\smc{l}}_{\omega_h ,
  \vect{p}}\right\}\; ,
\label{eq:2cont}
\end{multline}
and the terms $\Theta_{\omega_h , \vect{p}}$ and $\Lambda_{\omega_h ,
\vect{p}}$ are respectively given by:
\begin{equation}
\begin{split}
  \Theta_{\omega_h , \vect{p}} &= \Frac{1}{\nu L^2}\Frac{g^2
  \left[\omega(\vect{p}) - \mu\right] + g_c \left[\omega(\vect{p}) -
  \mu\right]^2 + g_c \omega_h^2}{\left\{g^2 + g_c
  \left[\omega(\vect{p}) - \mu\right]\right\}^2 + g_c^2 \omega_h^2} \\
  \Lambda_{\omega_h , \vect{p}}&= \Frac{1}{\nu L^2}\Frac{g^2
  \omega_h}{\left\{g^2 + g_c \left[\omega(\vect{p}) -
  \mu\right]\right\}^2 + g_c^2 \omega_h^2}\; .
\end{split}
\label{eq:thlam}
\end{equation}
Note that the the term $1/\nu L^2$ is here necessary since, in the
thermodynamic limit, the Coulomb and photon coupling constant have to
be rescaled with the volume according to the
definitions~\eqref{eq:resca}.

Finally, the mean-field value of the total action
\begin{equation}
  S^{\smc{ag}} [|\Delta| , g_{\text{eff}}] = \beta (\omega_c - \mu)
  |\psi|^2 + \beta |\Sigma|^2/g_c + S[\bar{Q}] \; ,
\label{eq:mfabg}
\end{equation}
coincides with the Abrikosov and Gor'kov theory for a superconductor
with magnetic impurities (see, e.g.,~\cite{maki}).

\begin{acknowledgments}
  We are grateful to P.R. Eastham, J. Keeling, R. S. Moir,
  M.V. Mostovoy, R. Smith, and M.H. Szymanska for suggestions and
  useful discussions. One of us (FMM) would like to acknowledge the
  financial support of EPSRC (GR/R95951). This work is supported by
  the EU Network ``Photon mediated phenomena in semiconductor
  nanostructures'' HPRN-CT-2002-00298. The NHMFL is supported by the
  National Science Foundation, the state of Florida and the US
  Department of Energy.
\end{acknowledgments}

%


\begin{thebibliography}{99}
%

\bibitem{hopfield}
J. J. Hopfield, \emph{Phys. Rev} \textbf{112}, 1555 (1958).

\bibitem{keldysh} 
L. V. Keldysh, ``Macroscopic coherent states of excitons in
semiconductors''. In \emph{Bose Einstein Condensation} (Eds.~ Griffin,
A., Snoke, D. W., and Stringari, S.), 246. Cambridge University Press,
Cambridge (1995).

\bibitem{snoke_book}
S. A. Moskalenko and D. W. Snoke, \emph{Bose-Einstein Condensation of
  Excitons and Biexcitons}, Cambridge University Press, Cambridge
(2000). 

\bibitem{haug}
H. Haug and S. Koch, \emph{Quantum Theory of the Optical and
  Electronic Properties of Semiconductors}, World Scientific (1990).

\bibitem{weisbuch} 
C. Weisbuch, M. Nishioka, A. Ishikawa, and Y. Arakawa,
\emph{Phys. Rev. Lett.} \textbf{69}, 3314 (1992).

\bibitem{lesidang}
Le Si Dang, D. Heger, R. Andr\'e, F. Boeuf, and R. Romestain,
\emph{Phys. Rev. Lett.} \textbf{81}, 3920 (1998).

\bibitem{lidzey}
D. G. Lidzey, D.~D.~C. Bradley, M.~S. Skolnick, T.~Virgili, S.~Walker,
and D.~M. Whittaker, \emph{Nature} \textbf{395}, 53 (1998).

\bibitem{savvidis_baumberg_saba}
P. G. Savvidis, J. J. Baumberg, R. M. Stevenson, M. S. Skolnick,
D. M. Whittaker, and J. S. Roberts, \emph{Phys. Rev. Lett.}
\textbf{84}, 1547 (2000); J. J. Baumberg, P. G. Savvidis,
R. M. Stevenson, A. I. Tartakovskii, M. S. Skolnick, D. M. Whittaker,
and J. S. Roberts, \emph{Phys. Rev. B} \textbf{62}, R16247 (2000);
M. Saba, C. Ciuti, J. Bloch, V. Thierry-Mieg, R. Andr\'e, Le Si Dang,
S. Kundermann, A. Mura, G. Bongiovanni, J. L. Staehli, and B. Deveaud,
\emph{Nature} \textbf{414}, 731 (2001).

\bibitem{huang} 
R. Huang, Y. Yamamoto, R. Andr\'e, J. Bleuse, M. Muller, and
H. Ulmer-Tuffigo, \emph{Phys. Rev. B} \textbf{65}, 165314 (2002).

\bibitem{deng} 
H. Deng, G. Weihs, C. Santori, J. Bloch, and Y. Yamamoto,
\emph{Science} \textbf{298}, 199 (2002); H. Deng, G. Weihs, D. Snoke,
J. Bloch, and Y. Yamamoto, \emph{PNAS} \textbf{100}, 15318 (2003).

\bibitem{tassone}
F. Tassone, C. Piermarocchi, V. Savona, A. Quattropani, and
P. Schwendimann, \emph{Phys. Rev. B} \textbf{56}, 7554 (1997);
F. Tassone and Y. Yamamoto, \emph{Phys. Rev. B} \textbf{59}, 10830
(1999).

\bibitem{tartakovskii} 
A. I. Tartakovskii, M. Emam-Ismail, R. M. Stevenson, M. S. Skolnick,
V. N. Astratov, D. M. Whittaker, J. J. Baumberg, and J. S. Roberts,
\emph{Phys. Rev. B} \textbf{62}, R2283 (2000).

\bibitem{pol_dynamics} 
See, e.g., A. Kavokin and G. Malpuech, \emph{Thin Films and
Nanostructures} Vol. \textbf{32}, \emph{Cavity Polaritons}, Academic
Press, Amsterdam (2003), and reference therein; D. Porras, C. Ciuti,
J. J. Baumberg, and C. Tejedor, \emph{Phys. Rev. B} \textbf{66},
085304 (2002).


\bibitem{keldysh_kopaev_kozlov} 
L. V. Keldysh and Yu. V. Kopaev, \emph{Fiz. Tverd. Tela} (Leningrad)
\textbf{6}, 2791 (1964) [\emph{Sov. Phys. Solid State} \textbf{6},
2219 (1965)]; 
L. V. Keldysh and A. N. Kozlov, \emph{Zh. \'Eksp. Teor. Fiz.}
\textbf{54}, 978 (1968) [\emph{Sov. Phys. JETP} \textbf{27}, 521
(1968)].

\bibitem{comte_nozieres}
C. Comte and P. Nozi\`eres, \emph{J. Physique} \textbf{43}, 1069
(1982); P. Nozi\`eres and C. Comte, \emph{J. Physique} \textbf{43},
1083 (1982).

\bibitem{stark}
 S. Schmitt-Rink and D. S. Chemla, \emph{Phys. Rev. Lett.} \textbf{57},
2752 (1986); C. Comte and G. Mahler, \emph{Phys. Rev B} \textbf{34},
7164 (1986).

\bibitem{laser_book}
M. O. Scully and M. S. Zubairy, \emph{Quantum Optics}, Cambridge
University Press, Cambridge (1997).

\bibitem{paul}
P. R. Eastham and P. B. Littlewood, \emph{Solid Sate Commun.}
\textbf{116}, 357 (2000); P. R. Eastham and P. B. Littlewood,
\emph{Phys. Rev. B} \textbf{64}, 235101 (2001).


\bibitem{marzena}
M. H. Szymanska and P. B. Littlewood, \emph{Solid Sate Commun.}
\textbf{124}, 103 (2002); M. H. Szymanska, P. B. Littlewood, and
B. D. Simons, \emph{ Phys. Rev. A} \textbf{68}, 013818 (2003).

\bibitem{zittartz}
J. Zittartz, \emph{Phys. Rev.} \textbf{164}, 575 (1967).

\bibitem{efetov}
K. B. Efetov, \emph{Supersymmetry in Disorder and Chaos}. Cambridge
University Press, Cambridge (1997).

\bibitem{sclaser} 
G.~P. Agrawal and N.~K. Dutta, \emph{Semiconductor Lasers},
2$^\text{nd}$ ed., Van Nostrand Reinhold, New York, (1993).



\bibitem{edwards_anderson}
S. F. Edwards and P. W. Anderson, \emph{J. Phys. F} \textbf{5}, 965
(1975).

\bibitem{damian}
 A. Altland, B. D. Simons, and D. Taras-Semchuk, \emph{Advances in
 Physics} \textbf{49}, 321 (2000).

\bibitem{lamacraft} 
A. Lamacraft and B. D. Simons, \emph{Phys. Rev. Lett.} \textbf{85},
4783 (2000); A. Lamacraft and B. D. Simons, \emph{Phys. Rev. B}
\textbf{64}, 014514 (2001).

\bibitem{usadel}
K. D. Usadel, \emph{Phys.~Rev. B} \textbf{4}, 99 (1971).

\bibitem{abrikosov_gorkov}
A. A. Abrikosov and L. P. Gor'kov, \emph{Sov.~Phys. JETP} \textbf{12},
1243 (1961).

\bibitem{maki}
K. Maki, ``Gapless superconductivity''. In \emph{Superconductivity}
(Ed.~Parks, R. D.), vol.~\textbf{2}, 1035. Marcel Dekker, Inc., New
York (1969).

\bibitem{keeling} 
J. Keeling, L. S. Levitov, and P. B. Littlewood,
\emph{Phys. Rev. Lett.} \textbf{92}, 176402 (2004).


\bibitem{eckern_pelzer}
U. Eckern and F. Pelzer, \emph{J. Low Temp. Phys.} \textbf{73}, 433
(1988).


\bibitem{smith_ambegaokar}
R. A. Smith and V. Ambegaokar, \emph{Phys. Rev. B} \textbf{62}, 5913
(2000).

\bibitem{moir} 
For a similar expansion in a different context, see R. S. Moir,
\emph{Collective Phenomenon in Itinerant Magnetism and
Superconductivity: Disorder and Competition}, PhD Thesis, Cambridge
(2003).

\bibitem{comment} 
Because we have used a low frequency expansion, the estimate of the
amplitude mode frequency, while correctly proportional to $|\Delta|$,
is numerically inaccurate: the exact value for a gapless BCS theory is
$2 |\Delta|$ [see P. B. Littlewood and C. M. Varma, \emph{Phys. Rev. B}
\textbf{26} 4883 (1982)]. The gapless phase mode is correctly
determined, however.


\bibitem{chaikin_lubensky} 
See, e.g., P. M. Chaikin and T. C. Lubensky, \emph{Principles of
Condensed Matter Physics}, Cambridge University Press, Cambridge
(1995).

\bibitem{footnote_fluct} 
Formally, \eqref{eq:phpro} is evaluated by expressing
$\theta_{\vect{p}}$ in terms of the fields~\eqref{eq:fluct},
$\theta_{\vect{p}} = (\delta\Delta_{\vect{p}}^{\smc{t}} - \delta
\Gamma_{\vect{p}}^{\smc{t}})/2 g$ and, after integrating over the
fields $\delta \Gamma$, making use of the propagator~\eqref{eq:kerne}
for the fields $\delta \Delta$. The dominant contribution, which
results in the power-law dependence of~\eqref{eq:phpro}, derives from
the massless mode.

\bibitem{timmermans_holland_ohashi_griffin}
E. Timmermans, K. Furuya, P.W. Miloni, and A. K. Kerman,
\emph{Phys. Lett. A} \textbf{285}, 228 (2001); M. Holland,
S.J.J.M.F. Kokkelmans, M.L. Chiofalo, and R. Walser,
\emph{Phys. Rev. Lett.} \textbf{87}, 120406 (2001); Y. Ohashi and
A. Griffin, \emph{Phys. Rev. Lett.} \textbf{89}, 130402 (2002);
Y. Ohashi and A. Griffin, \emph{Phys. Rev. A} \textbf{67}, 063612
(2003).

\end{thebibliography}
\end{document}